\documentclass{aa}

\usepackage[varg]{txfonts}
\usepackage{amssymb}
\usepackage{upgreek}
\usepackage{graphicx}
\usepackage{epstopdf}
\usepackage{amsmath}
\usepackage{color}
\usepackage{natbib}
\usepackage{hyperref}
\usepackage{gensymb}

\usepackage{cleveref}

\usepackage{aas_macros}

\usepackage{bm}
\usepackage{mathtools}
\usepackage{lipsum}
\usepackage{physics}

\newcommand{\be}{\begin{equation}}
\newcommand{\ee}{\end{equation}}
\newcommand{\bc}{\begin{center}}
\newcommand{\ec}{\end{center}}
\newcommand{\beq}{\begin{eqnarray}}
\newcommand{\eeq}{\end{eqnarray}}

\newcommand{\df}[1]{\mathrm{d}{#1}}
\newcommand{\rmd}{{\rm d}}

\newcommand{\unit}[1]{\mbox{\boldmath $\hat{#1}$}}
\newcommand{\rs}{R_{\rm S}}

\bibpunct{(}{)}{;}{a}{}{,} 

\DeclareUnicodeCharacter{00A0}{ }

\makeatletter
\def\fvec#1{\underline{\sbox\tw@{$#1$}\dp\tw@\z@\box\tw@}}
\makeatother

\begin{document}

\title{Analytical techniques for polarimetric imaging of accretion flows in Schwarzschild metric}

\titlerunning{Accretion disc polarization in Schwarzschild metric}

\author{Vladislav~Loktev\inst{1}
\and  Alexandra~Veledina \inst{1,2,3}
\and  Juri~Poutanen\inst{1,2,3}}

\authorrunning{Loktev et al.} 

\institute{Department of Physics and Astronomy, FI-20014 University of Turku, Finland\\ \email{[vladislav.loktev,alexandra.veledina,juri.poutanen]@utu.fi}
   \and Nordita, KTH Royal Institute of Technology and Stockholm University,  SE-10691 Stockholm, Sweden
   \and Space Research Institute of the Russian Academy of Sciences, Profsoyuznaya str. 84/32, 117997 Moscow, Russia 
}

\date{Received XXXX / Accepted XXXX}

\abstract{Emission from an accretion disc around compact objects, such as neutron stars and black holes, is expected to be significantly polarized. 
The polarization can be used to put constraints on geometrical and physical parameters of the compact sources -- their radii, masses and spins -- as well as to determine the orbital parameters. 
The radiation escaping from the innermost parts of the disc is strongly affected by the gravitational field of the compact object and relativistic velocities of the matter.
The straightforward calculation of the observed polarization signatures involves computationally expensive ray-tracing technique.
At the same time, having fast computational routines for direct data fitting becomes increasingly important in light of the currently observed images of the accretion flow around supermassive black hole in M87 by the Event Horizon Telescope, infrared polarization signatures coming from Sgr~A*, as well as for the upcoming X-ray polarization measurements by the \textit{Imaging X-ray Polarimetry Explorer} and \textit{enhanced X-ray Timing and Polarimetry} mission.
In this work, we obtain an exact analytical expression for the rotation angle of polarization plane in Schwarzschild metric accounting for the effects of light bending and relativistic aberration. 
We show that the calculation of the observed flux, polarization degree and polarization angle as a function of energy can be performed analytically with high accuracy using approximate light-bending formula, lifting the need for the pre-computed tabular models in fitting routines.
}

 
\maketitle

\section{Introduction}\label{sec:intro}

Accretion discs are among the most efficient energy conversion engines in the Universe.
They surround the dense-most objects -- neutron stars (NSs) and black holes (BHs).
As the matter gradually shifts towards the compact object, it releases the excess of gravitational energy via radiation.
Physical mechanisms leading to energy liberation and geometrical properties of emitting medium are subject of study of the modern high-energy astrophysics.
Recent investigations have been focused on the innermost parts of the accretion disc, in the regime of strong gravity, where the matter predominantly radiates in the X-ray energies.
X-ray spectroscopy  \citep{Reynolds14,Bambi2021} and timing techniques \citep{Revnivtsev1999,GRM03,Uttley2014,Axelsson21} have been exploited to trace the geometry of the inner parts of the accretion flow and to identify contributions of various components emitting in the X-ray energies.

Polarimetry is known to be a fine measure of the geometry and radiative processes operating in the source.
Its full capacity in the X-ray range will be used with the upcoming launch of dedicated polarimetric satellites, such as \textit{Imaging X-ray Polarimetry Explorer} \citep[\textit{IXPE},][]{weisskopf16} and \textit{enhanced X-ray Timing and Polarimetry} mission \citep[\textit{eXTP},][]{Zhang2019eXTP}.
Numerous efforts are being aimed at predicting and finding distinct signatures of the accretion disc in strong gravity regime from the polarimetric information \citep[e.g.,][]{Dovciak2008,Li2009,Ingram15}. 
For the BH X-ray binaries, in the absence of a solid surface of the compact object, accretion is the only source of the observed X-ray emission, and the obtained signatures can be directly connected to the innermost geometry of the accretion disc.
Polarimetry in the infrared and millimetre bands was recently shown  to be a powerful tool to study the structure of accretion flows in the vicinity of supermassive BHs in the Milky Way \citep{Gravity18,Bower18} and in M87 \citep{EHT21a,EHT21b}. 
For the NS X-ray binaries, such as accreting millisecond pulsars, the disc can serve as a source of constant energy-dependent polarimetric background, which has to be subtracted to obtain polarimetric profiles of pulsations \citep{VP04}.

First studies of polarimetric signatures of the accretion disc have been presented in \citet{Rees1975}, who considered the flat space and dominant role of electron scattering, thus applying earlier results on the electron-scattering stellar atmospheres \citep{Cha60,Sob63} to the disc in the Newtonian approximation.
Observing such polarization signatures has been considered crucial for confirming the existence of an accretion disc in the first place \citep{LightmanShapiro1975}. 
A number of deviations from this simple model were discussed, such as the role of true absorption (Monte-Carlo estimates for this case have been presented in \citealt{LightmanShapiro1975}; analytic results for such atmospheres are discussed in \citealt{Losob79,LoskutovSobolev1981}) and \citet{BardeenPatterson1975} alignment of the inner parts of the disc with the BH spin.

Yet another important addition to the early works on the disc polarization signatures is inclusion of the effects of special and general relativity \citep{ConnorsStark1977,StarkConnors1977,PineaultRoeder1977kerr_analyt,PineaultRoeder1977kerr_num}.
Relativistic aberration and light deflection lead to a rotation of the polarization angle (PA) and alter the viewing angle of different segments of the disc, the latter effect leading to a different polarization degree (PD) of a given segment; while the PD is Lorentz invariant, the difference with respect to the non-relativistic case appears because of the different angle between the disc normal and observers' direction.
Frame dragging effects, relevant to spinning BHs described by \citet{Kerr1963} metric, lead to additional rotation of polarization plane along the ray.
Calculations of all these effects involves parallel transport of the polarization vector along null geodesics, which needs calculations of \citet{WalkerPenrose1970} constant of motion.
For every null geodesics leaving the disc, the polarization at infinity can be evaluated by solving linear equations for the components of polarization vector \citep[e.g.,][]{ConnorsStark1977,Connors1980,Dovciak2008,Li2009,Ingram15}.

In this implicit formulation, the calculation of Stokes parameters is related to the computationally expensive numerical integration of the equation of geodesic, known as the ray-tracing technique.
With the launch of X-ray polarimetric satellites, the data fitting procedures will need fast routines for calculating polarization from accretion discs.
High time gain can be achieved, e.g., by tabulating the \textit{observed} polarization characteristics of the disc rings for different BH and orbit parameters, so that the minimization routines would only need to proceed via interpolation between the pre-computed models.
On the other hand, the tabulated models would have to be recalculated with any change to the local model, such as alteration of the angular dependence of emission or radial energy dissipation profile. 
Similar problems are faced with in the analysis and theoretical modelling of polarimetric images of accretion flows around supermassive BHs. 
To accelerate the calculations, \citet{Narayan21} recently developed a fast code for evaluating polarimetric images in Schwarzschild metric, which is based on evaluation of Walker-Penrose constants and approximate light bending formula of \citet{B02}.
Understanding the role of different relativistic effects and their separation in the total image, however, remains elusive, as explicit analytical expressions for the PA and PD transformation --- as a result of the joint action of general and special relativity (GR and SR) --- have not been presented so far.

Using such explicit formulae to directly relate the local and the observed PA and PD can also substantially speed up the minimisation routine. 
\citet{Connors1980} give explicit analytical expression for the rotation angle of polarization plane due to solely SR effects (their eq.~18).
\citet{Pineault1977pol_schw} pointed out that total rotation of PA in Schwarzschild metric and after accounting for relativistic motions in the disc is not a simple sum of rotations caused by the relativistic motion in the flat space ($\chi^{\rm SR}_{\rm flat}$) and the light bending ($\chi^{\rm GR}$).

In this work, we derive explicit analytical expressions for the rotation of the PA accounting for relativistic motion of matter in the accretion disc and light bending in Schwarzschild metric.  
Corresponding formulae were previously derived for polarization properties of rapidly rotating NSs \citep{VP04,Poutanen2020polarization,Loktev20}.
We use the laws of geometrical optics and exploit the fact that the light trajectories are flat in Schwarzschild metric, i.e. the orientation of the polarization vector is fixed with respect to the angular momentum of the ray trajectory \citep[see][]{Pineault1977pol_schw}.
The light bending is computed using the recent analytical approximation derived in \citet{Poutanen2020bending}, which allows us to achieve high accuracy, with the deviations from the exact solution being smaller than the existing or expected statistical errors of observations.
This removes the need for using computationally expensive ray-tracing algorithms when computing polarization of the accretion discs around NSs or low-spin BHs.

\section{Polarized radiation from the accretion disc}

\subsection{Observed flux}

{We consider emission of a disc surface element in an axially symmetric, geometrically (infinitely) thin accretion disc in Schwarzschild metric.}
We compute the observed Stokes parameters following the approach described in \citet{Poutanen2020bending} and repeat here the basic notations and formulae for completeness. 

\begin{figure}
\centering
\includegraphics[width=0.95\columnwidth]{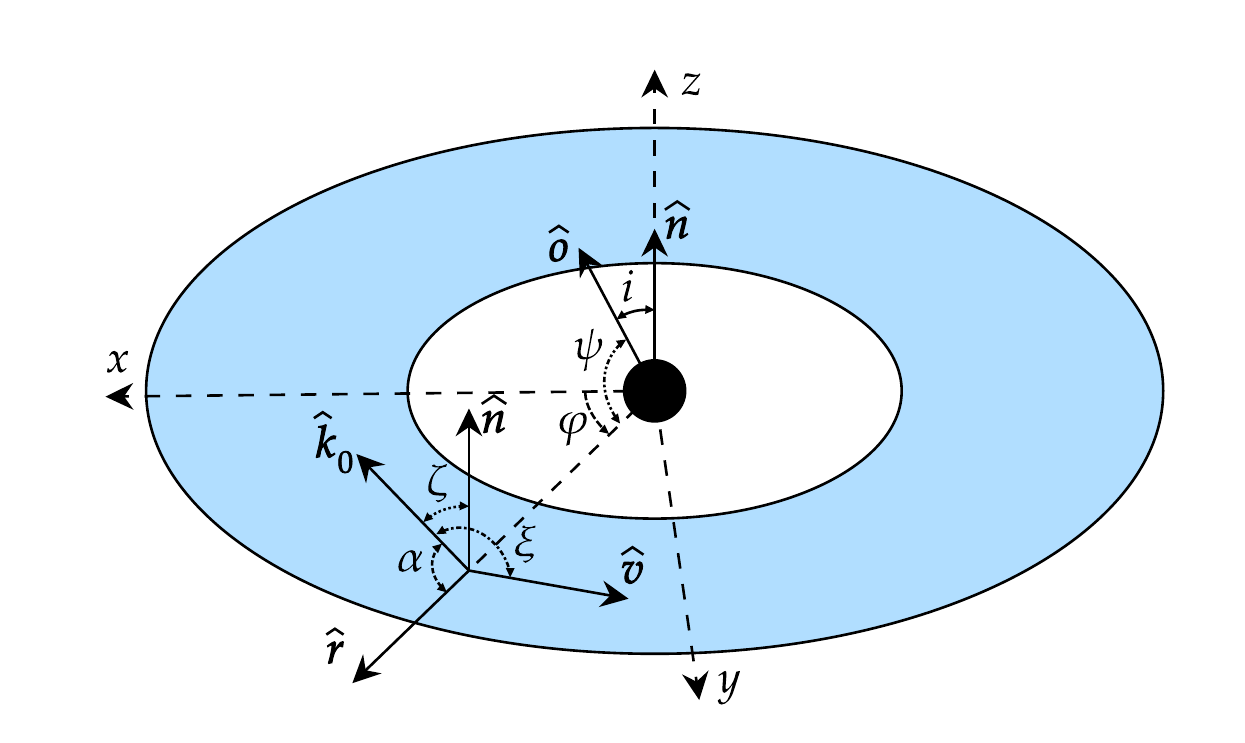} 
\caption{Considered geometry of a flat accretion disc ring. The emitting surface element is described by the radius-vector $\bm{r}$ and  velocity $\bm{v}$. 
Vector $\unit{o}$ pointing in the observer direction makes angle $i$ with the normal.
}
\label{fig:geom}
\end{figure}

We choose a Cartesian coordinate system with the $z$-axis coinciding with the normal to the disc and $x$-axis lying along the projection of the line of sight on the disc (see Fig.~\ref{fig:geom}).
In this system, the normal to the disc $\unit{n}$, the unit vector in the observer direction $\unit{o}$ and the radius-vector of the surface element $\unit{r}$ have the following coordinates: 
\begin{eqnarray}\label{eq:geom_vectors}
\unit{n} & = & (0,0,1),\nonumber \\
\unit{o} & = & (\sin i,0,\cos i), \\   
\unit{r} & = & (\cos\varphi,\sin\varphi,0),\nonumber 
\end{eqnarray}
where $\varphi$ is the azimuth of the radius-vector measured from the $x$-axis and $i$ is the disc inclination to the line of sight.
The radius-vector of the surface element makes angle $\psi$ to the line of sight:
\begin{equation}\label{eq:cospsi}
 \cos\psi=\unit{r} \cdot \unit{o} =  \sin i\ \cos\varphi.
\end{equation}

The photon trajectories are planar in Schwarzschild metric, hence the direction of the photon momentum close to the disc surface can be described as a linear combination of the observer vector and the radius-vector of the emission point:
\begin{equation}\label{eq:k0}
\unit{k}_0=[ \sin\alpha\ \unit{o} +\sin(\psi-\alpha)\ \unit{r}]/\sin\psi, 
\end{equation}
where 
\begin{equation}\label{eq:cosalpha}
\cos\alpha=\unit{r}\cdot \unit{k}_0.
\end{equation}
The relation between the angles $\alpha$ and $\psi$ is described by the light bending integral or can be approximated using a simple analytical formula \citep{PFC83,B02,PB06,SNP18,Poutanen2020bending}. 

We consider purely Keplerian motion of disc matter, with the velocity unit vector being parallel to the azimuthal vector
\begin{equation}\label{eq:circvel} 
\unit{v}= \unit{\varphi} \equiv  (-\sin\varphi,\cos\varphi,0). 
\end{equation}
The dimensionless velocity $\beta=v/c$ relative to a static observer at the circumferential radius $r=R/\rs$, measured in units of the Schwarzschild radius $\rs=2GM/c^2$ of the central object of mass $M$, is \citep[see e.g.][]{Luminet1979}
\begin{equation}\label{eq:beta} 
\beta = \sqrt{\frac{u}{2(1-u)}}, 
\end{equation}
where $u=1/r$ is the compactness. 
The corresponding Lorentz factor is 
\begin{equation}\label{eq:gamma}
\gamma=\frac{1}{\sqrt{1-\beta^2}}= \sqrt{ \frac{1-u}{1-3u/2}} .
\end{equation}
The photon momentum makes angle $\xi$ with the velocity vector,
\begin{equation}\label{eq:cosxi2}
\cos\xi=\unit{v} \cdot \unit{k}_0
=\frac{\sin\alpha}{\sin\psi} \unit{v} \cdot \unit{o}=
- \frac{\sin\alpha}{\sin\psi}\sin i\ \sin\varphi\ ,
\end{equation}
and angle $\zeta$ with the disc normal,
\begin{equation}\label{eq:coszeta}
  \cos\zeta =\unit{n} \cdot \unit{k}_0 = \frac{\sin\alpha}{\sin\psi} \unit{n} \cdot \unit{o} =  \frac{\sin\alpha}{\sin\psi} \cos i  . 
\end{equation}
The Doppler factor is 
\begin{equation}\label{eq:dop}
\delta=\frac{1}{\gamma(1-  \beta\, \unit{v} \cdot \unit{k}_0)} = \frac{1}{\gamma(1-\beta\cos\xi)} .
\end{equation}
The unit vector of the photon momentum in the frame comoving with the surface element (fluid frame hereafter) is computed using Lorentz transformation: 
\begin{equation} \label{eq:k0Lorentz}
\unit{k}^{\prime}_0 = \delta \left[ \unit{k}_0 - \gamma\beta \unit{v} + (\gamma-1) \unit{v} (\unit{v}\cdot \unit{k}_0 ) \right] .
\end{equation}
From this, we get the angle between the photon momentum and the local normal in the fluid frame 
\begin{equation}\label{eq:cosalphadop}
\cos\zeta'= \delta\cos\zeta .
\end{equation}
The projection of the photon momentum on the disc plane in that frame has azimuth $\phi'$ (as measured from the radial direction) and can be determined through 
\begin{eqnarray} \label{eq:sinphiprime}
\sin{\phi'} & = & \frac{\unit{k}^{\prime}_0 \cdot \unit{\varphi}} {\sin\zeta'} 
= \delta\gamma\  \frac{\cos \xi   - \beta} {\sin\zeta'} 
, \\ 
\label{eq:cosphiprime} 
\cos{\phi'} & = & \frac{\unit{k}^{\prime}_0 \cdot \unit{r} } {\sin\zeta'} = \delta\ \frac{\cos\alpha} {\sin\zeta'}.
\end{eqnarray}
Specific flux observed from the surface element at photon energy $E$ is  
\begin{equation} \label{eq:dF_E}
 \rmd F_E=I_E\ \rmd\Omega,
\end{equation} 
where 
$\df{\Omega}$ is the solid angle occupied by the surface element on the observer's sky, 
$I_E$ is the specific intensity of radiation at infinity, which is related to that in the fluid frame 
\begin{equation}
  I_{E} = \left (\frac{E}{E'}\right )^3 I'_{E'} (\zeta') . 
\end{equation} 
Here we assume that $I'$ does not depend on the azimuthal angle $\phi'$ and for brevity omit its dependence on radius $r$.  
The total redshift factor \citep{Luminet1979,Chen89},
\begin{equation} \label{eq:EEpr}
g \equiv \frac{E}{E'}  = \delta \sqrt{1-u} =  \frac{\sqrt{1-3u/2} } {1+\beta\sin i\sin\varphi\sin\alpha/\sin\psi } ,
\end{equation}
combines the effects of the gravitational redshift and the Doppler effect. 
The solid angle occupied by the surface element of area $\rmd S=R \rmd R \rmd\varphi/\sqrt{1-u}$ is given by 
\begin{equation}\label{eq:omega_disc}
\rmd \Omega=\frac{\rmd S \cos \zeta}{D^2} {\cal D} = 
\frac{\rs^2}{D^2}\ \frac{r\, \rmd r\ \rmd\varphi}{\sqrt{1-u}}  \ {\cal D} \ \cos \zeta.
\end{equation}
Here $\rmd S \cos \zeta$ is the projection of the element area on the plane of the sky, $D$ is the distance to the source, and
\begin{equation}\label{eq:lensing}
{\cal D} = 
\frac{1}{1-u} \frac{\rmd\cos\alpha}{\rmd\cos\psi}  
\end{equation}
is the lensing factor \citep{B02,Poutanen2020bending}. 
We obtain the final expression for the observed spectral flux in the form
\begin{equation}\label{eq:fluxspot}
\df{F}_{E} (r,\varphi) =  g^{3} I'_{E'}(\zeta') \ 
\frac{\df{S}  \cos\zeta}{D^2} \ {\cal D}. 
\end{equation}

The total observed flux from the disc can then be obtained by integrating Eq.~(\ref{eq:fluxspot}) over the radius and azimuthal angle
\begin{equation}\label{eq:flux_azimuth_ave}
F_E =  \frac{\rs^2}{D^2}  
 \int\limits_{r_\textrm{in}}^{r_\textrm{out}}   \frac{r\, \rmd r}{\sqrt{1-u}} \int_0^{2\uppi}  \rmd \varphi  \, 
 g^{3}\ {\cal D}\ \cos\zeta\  I'_{E'}(\zeta') . 
\end{equation}
For a chosen inclination $i$ and for every pair $(r,\varphi)$ we compute $\psi$ using Eq.~(\ref{eq:cospsi}), which is then used to compute $\alpha$  and ${\cal D}$ --- either exactly from elliptical integrals (as described, e.g. in \citealt{SNP18}), or using analytical formulae (see \citealt{Poutanen2020bending} and Sect.\,\ref{sec:bending}).
Then, $\xi$ and $\zeta$ are obtained from Eqs.~(\ref{eq:cosxi2}) and (\ref{eq:coszeta}), respectively.
Using the Keplerian velocity and the Lorentz factor given by Eqs.~\eqref{eq:beta} and \eqref{eq:gamma}, we then get the Doppler factor $\delta$ from Eq.~\eqref{eq:dop}. 
Further, from Eqs.~\eqref{eq:EEpr} and \eqref{eq:cosalphadop}, we get the photon the energy $E'$ and the zenith angle $\zeta'$ in the fluid frame, which are needed to obtain $I'_{E'}(\zeta')$, and if needed the azimuth $\phi'$ from Eqs.~\eqref{eq:sinphiprime} and \eqref{eq:cosphiprime}. 

\subsection{Computing polarized flux}

We consider only linear polarization, which is described by the three-component Stokes vector.\footnote{The fourth component describing circular polarization is affected by light bending and aberration exactly the same way as the intensity $I$ and the circular polarization degree is conserved along photon trajectory in the absence of plasma effects.}
The direction of the polarization vector changes along the photon trajectory because of the gravitational light bending and aberration.
The key property of Schwarzschild metric allowing us to derive simple analytical formulae for the PA rotation is that the angle between the polarization vector and the trajectory plane is conserved.

In order to describe polarized radiation in the fluid frame, it is convenient to introduce the polarization basis formed by the vector of the local normal and the photon momentum $\unit{k}_0'$: 
\be\label{eq:polbas_k0n}
\unit{e}_1'^{0} = \frac{\unit{n}-\cos{\zeta'}\  \unit{k}_0'}{\sin{\zeta'}},\qquad 
\unit{e}_2'^{0} = \frac{\unit{k}_0' \times \unit{n}}{\sin{\zeta'}} .
\ee
Radiation field emitted at angle $\zeta'$ from the disc surface with coordinates $(r,\varphi)$ is then described by the Stokes vector  
\begin{equation} \label{eq:primepolvec}
\bm{I}'_{E'}(\zeta')=  {I}'_{E'}(\zeta')
\begin{bmatrix} 1\\ p(\zeta') \cos2\chi_{0} \\ p(\zeta') \sin2\chi_{0} \end{bmatrix} ,
\end{equation}
where $p$ is the linear PD of radiation, which is invariant, i.e. does not change along photon trajectory.
The PA $\chi_{0}$ is defined as the angle between the polarization vector and the basis vector $\unit{e}_1'^{0}$ measured in the counterclockwise direction. 
Transformation of vector $\bm{I}'_{E'}(\zeta')$ to the observed Stokes vector involves rotation by the corresponding Mueller matrix 
\begin{equation}\label{eq:pol-rot-matr}
    {\bf M}(r,\varphi) = \begin{bmatrix}
        1 & 0 & 0 \\
        0 & \cos2\chi^{\rm tot} & -\sin2\chi^{\rm tot} \\
        0 & \sin2\chi^{\rm tot} & \cos2\chi^{\rm tot} 
    \end{bmatrix} , 
\end{equation}
where $\chi^{\rm tot}$ is the rotation angle of polarization plane. 
Combining this matrix with Eq.~(\ref{eq:fluxspot}), we get the expression for the observed Stokes vector (in terms of fluxes):  
\begin{equation}\label{eq:Stokes_spot}
\df{\bm{F}}_{E} (r,\varphi) =  g^{3}\ {\bf M}(r,\varphi)\ \bm{I}'_{E'}(\zeta')\ 
 \frac{\df{S}  \cos\zeta }{D^2}\  {\cal D}. 
\end{equation}
Integrating over azimuth and radius, we get the observed Stokes vector from the entire disc surface: 
\begin{equation} \label{eq:polflux}
 \bm{F}_E \equiv \begin{bmatrix} F_I\\ F_Q \\ F_U \end{bmatrix} = \frac{\rs^2}{D^2} \!\! 
 \int\limits_{r_\textrm{in}}^{r_\textrm{out}} 
 \!\!\! \frac{r\, \rmd r}{\sqrt{1-u}}  \int_0^{2\uppi}\!\!\!\!\!\!  \rmd \varphi  \, g^{3} {\cal D} \cos\zeta\  {\bf M}(r,\varphi)\bm{I}'_{E'}(\zeta') . 
\end{equation}

If the angular distribution of radiation escaping from the disc surface (as measured in the fluid frame) does not depend on the azimuth $\phi'$, but only on the zenith angle $\zeta'$, the polarization vector is parallel to one of the basis vectors \eqref{eq:polbas_k0n} and the Stokes $U$-parameter vanishes, i.e. $\chi_{0}=0$ or $\uppi/2$.
Equation~(\ref{eq:Stokes_spot}) transforms to
\begin{equation}\label{eq:Stokes_spot2}
\df{\bm{F}}_{E} (r,\varphi) =  g^{3}  {I}'_{E'}(r,\zeta') 
 \begin{bmatrix} 1\\ \pm p(\zeta') \cos2\chi^{\rm tot} \\ \pm p(\zeta') \sin2\chi^{\rm tot} \end{bmatrix}   
\ \frac{\df{S} \cos\zeta}{D^2}\ {\cal D}. 
\end{equation}
Integrating over $r$ and $\varphi$, we easily get the total observed Stokes vector.  
We note that the transformation of the Stokes vectors does not depend on the nature of radiation escaping from the disc or on the assumption about direction of the polarization vector in the fluid frame; it is fully determined by the rotation of polarization plane $\chi^{\rm tot}$.
In order to obtain an analytical solution relating the emitted and observed Stokes vectors, we need to find an explicit formula for $\chi^{\rm tot}$.

\subsection{Polarization angle}
  
Let us consider physical reasons for the rotation of polarization plane.
The electric field of the emitted photon oscillates in the plane perpendicular to the direction of photon propagation.
The polarization angle (PA), as determined by the direction of oscillation of the electric vector; in the fluid frame it is measured with respect to the projection of the local normal $\unit{n}$ on the sky.
As the photon propagates to the observer, the polarization vector remains perpendicular to the direction of motion, and hence rotates around the normal to the trajectory plane. 
In a general case, when the observer is not located in the disc plane, the normal to the trajectory plane does not coincide with the disc normal $\unit{n}$, hence the angle between the polarization plane and $\unit{n}$ changes.
Accordingly, this causes changes of the PA.

The PA seen by the observer is the sum of the PA of emitted photon and its total rotation along the trajectory:
\begin{equation}\label{eq:xhi0chitot}
    \chi = \chi_0 + \chi^{\rm tot} .
\end{equation}
The PA rotation can be computed directly from the rotation of the polarization plane and the projection of the polarization vector onto the sky, as viewed by the observer.
The rotation by the general relativistic (GR) light bending effects is (see derivation in Appendix~\ref{sec:app_gr})
\begin{equation}\label{eq:tanGR}
\tan\chi^{\text{GR}}  
= \frac{a\cos i \sin\varphi}{\sin i + a\cos\varphi}, 
\end{equation}
where 
\begin{equation}
a
= \frac{\cos\alpha - \cos\psi}{1- \cos\alpha  \cos\psi} .
\end{equation} 
This expression accounts only for GR (light bending) effects, while in reality the matter in the disc moves around the compact object at relativistic velocities, hence we need to account also for the SR effects, namely, for the relativistic aberration.

\begin{figure*}
\centering
\includegraphics[width=0.95\textwidth]{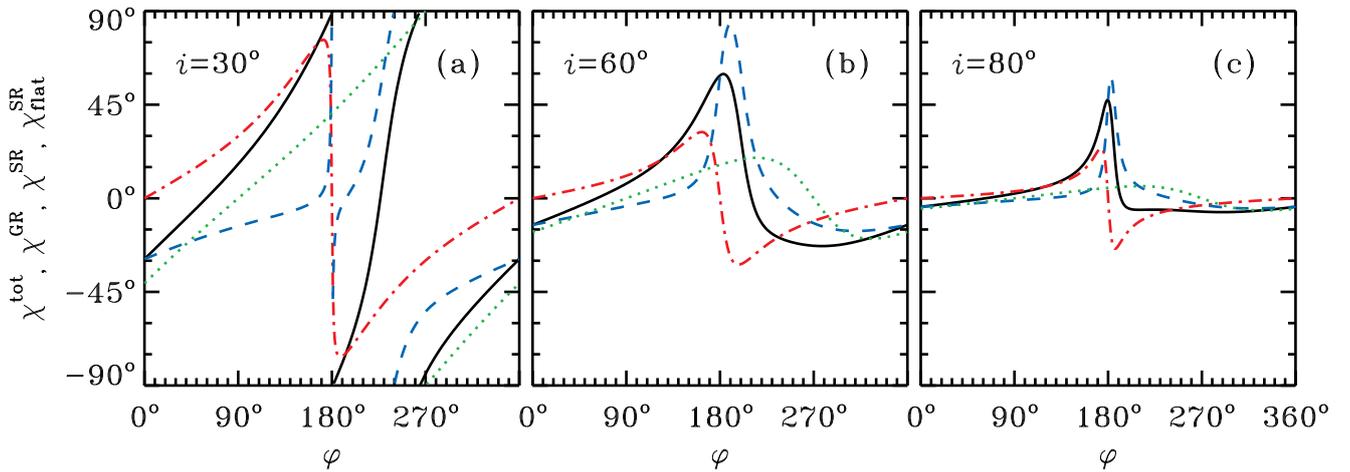}
\caption{
The rotation angles of polarization plane $\chi^{\mathrm{GR}}$ (red dot-dashed lines), $\chi^{\rm SR}$ (blue dashed lines), $\chi^{\rm SR}_{\rm flat}$ (green dotted lines) and $\chi^\text{tot}$ (black solid lines) for a ring at $r=3$ at different viewing angles (\textit{panel a}) $i=30\degr$, $60\degr$ (\textit{b}) and $i=80\degr$ (\textit{c}). 
}
\label{fig:chis2}
\end{figure*}

If the surface element moves at velocity $\bm{\beta} = \beta\unit{v}$, the photon vector $\unit{k}_0$, as measured in the observers (static) frame, is related to the vector $\unit{k}_0'$, as measured in the fluid frame, via the Lorentz transformation.
This transformation leads to the additional rotation of the PA, which can be written as (see Appendix~\ref{sec:app_sr} for derivation)
\begin{equation}\label{eq:tanSRGR}
 \tan\chi^{\text{SR}} 
 =   -\beta\ \frac{\cos \alpha \cos\zeta}{\sin^2\zeta-\beta \cos\xi}.
\end{equation}
We note that this expression takes into account the fact that the photon reaching the observer experiences light bending, and hence the outgoing photon in the static reference frame close to the disc surface propagates along vector $\unit{k}_0$, not $\unit{o}$. 

For the case of a flat space, when $\alpha=\psi$, the expression above reduces to 
\begin{equation}\label{eq:tanSRflat}
\tan\chi^{\text{SR}}_{\rm flat} = -\beta\ \frac{\cos i \cos\varphi}
    {\sin i +  \beta\sin\varphi },
\end{equation}
which is equivalent to eq.~(18) of \citet{Connors1980} for the case of the disc polarization perpendicular to meridional plane made by the photon momentum and the local normal, i.e. $\chi_0=\uppi/2$.

The total rotation of the PA is the sum of two rotations (Eqs.~\ref{eq:tanGR} and \ref{eq:tanSRGR}):
\begin{equation}\label{eq:chitot}
    \chi^\text{tot} =\chi^{\text{SR}} + \chi^{\text{GR}}.
\end{equation}
This expression is different from the simple addition of rotations due to light bending $\chi^{\rm GR}$ and rotation in flat space $\chi^{\rm SR}_{\rm flat}$ (Eq.~\ref{eq:tanSRflat}), as originally noted in \citet{Pineault1977pol_schw}.
The reason for this is that the Lorentz transformation has to be applied to the photon momentum already accounting for the bending effect.

\subsection{Light bending} 
\label{sec:bending}

The formulae presented above allow us to directly transform the Stokes parameters of radiation escaping from the surface element to the observed ones. 
To obtain the complete analytical transformation, we need to have an expression for $\alpha(\psi,r)$.
Exact integral equation for the inverse function, $\psi(\alpha,r)$ \citep{PFC83,B02,Poutanen2020bending}, can be used to compute the original expression at high accuracy.
For this, the function $\psi(\alpha,r)$ has to be tabulated at a dense grid of arguments and then the function $\alpha(\psi,r)$ can be obtained by interpolation \citep{SNP18}.

Alternatively, an approximate analytical expression \citep{B02} can be used: 
\begin{equation}\label{eq:b02}
  \cos\alpha \approx 1 - (1-u) y, 
\end{equation}
where $y = 1-\cos\psi$. 
This approximation gives ${\cal D}=1$ for the lensing factor \eqref{eq:lensing}. 
This relation was recently used in \citet{Narayan21} to compute polarized images of an accretion disc  in Schwarzschild metric observed nearly face-on. 
The approximate relation (\ref{eq:b02}) is accurate only when $\psi \lesssim 90\degr$ and therefore cannot be used for high inclinations.  

Higher accuracy can be achieved by using the recently proposed relation \citep{Poutanen2020bending}:
\begin{equation}\label{eq:bend_pout}
    \cos\alpha \approx 1 - y(1-u) \left\{1+\frac{u^2y^2}{112}-\frac{euy}{100} \left[\ln(1-\frac{y}{2})+\frac{y}{2}\right]\right\},
\end{equation}
where $e$ is the base of natural logarithm.
This relation can also be used for high inclinations.
In particular, this relation gives relative accuracy in $\alpha$  better than 0.06\% at $\psi < 120\degr$ and any radius exceeding 1.5$\rs$. 
At these radii, the error does not exceed 0.2\% for $\psi < 162\degr$ (which is valid for any azimuthal angle $\varphi$ and inclinations $i<72\degr$).
Using Eq.~\eqref{eq:bend_pout}, we get a similarly accurate analytical representation for the lensing factor: 
\begin{equation}\label{eq:lensing_pout}
{\cal D} \approx  1 + \frac{3u^2y^2}{112}  - \frac{e}{100} u y \left[ 2\, \ln\left(1-\frac{y}{2}\right) + y\frac{1-3y/4}{1-y/2}\right] .
\end{equation} 
Equations \eqref{eq:bend_pout} and \eqref{eq:lensing_pout} can be used instead of the exact relations for fast and accurate calculations of the observed Stokes parameters, as well as polarized images.

\subsection{Images} 
\label{sec:image}

The formalism developed above allows us to obtain images of the accretion disc on the sky in polarized light. 
For any point on the disc with polar coordinates $(r,\varphi)$, we first compute the angle between the observer direction and the radius vector of the element $\psi$ using Eq.~\eqref{eq:cospsi}. 
Then, using either exact or approximate relations for light bending, we get angle $\alpha$, which is then used to get the impact parameter \citep{PFC83,B02} 
\begin{equation}
b= \frac{r}{\sqrt{1-u}} \sin\alpha 
\end{equation}
in units of $\rs$. 
Owing to the fact that photon trajectories are planar is Schwarzschild metric, we get the position angle $\Phi$ (measured counterclockwise from the projection of the disc axis on the sky) of the point where photon hits the plane of the sky:   
\begin{eqnarray}
\sin\Phi & =&  - \frac{\sin\varphi}{\sin\psi} , \\
\cos\Phi & =&  - \frac{\cos i \cos\varphi}{\sin\psi} . 
\end{eqnarray} 
The error on $b$ is fully determined by the accuracy of the relation $\alpha(\psi)$, while $\Phi$ is exact.

\begin{figure*}
\centering
\includegraphics[width=0.95\textwidth]{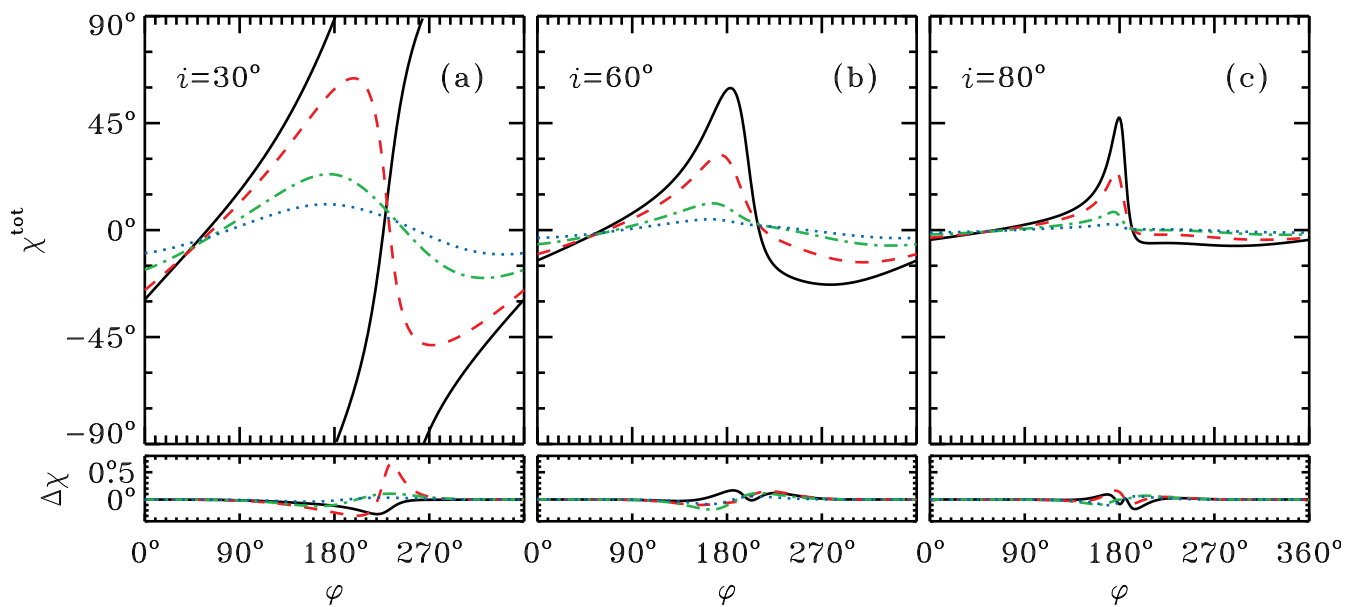}
\caption{\textit{Upper panels}: PA $\chi^{\rm tot}$  as a function of azimuth for disc inclinations $i=30\degr$ (\textit{panel a}), $60\degr$ (\textit{b}) and $80\degr$ (\textit{c}) 
and different $r= 3$ (black solid line), 5 (red dashed), 15 (green dot-dashed) and 50 (blue dotted). 
\textit{Lower panels}: Difference in the PA rotation computed using analytical formula \eqref{eq:bend_pout} for light bending and via  exact calculations.}
\label{fig:chis}
\end{figure*}

\section{Applications}

In this section we show the examples of calculations of the observed polarization signatures affected by the GR and SR effects.
We start with the simple decomposition of the action of these effects on the PA rotation ($\chi^{\rm GR}$ and $\chi^{\rm SR}$).
We then proceed to comparison between PA rotations obtained using the exact and approximate $\alpha(\psi)$ relations.
This relation is also used when obtaining the angle of the outgoing photon to the disc normal, which ultimately affects the observed flux and PD.
Finally, we show the calculations of the polarization signatures of the geometrically thin, optically thick \citep{SS73,NT73} accretion disc.

\subsection{Stokes vector for a narrow ring}

Relativistic effects on polarization are expected to be of highest importance at smallest disc radii.
First, we consider radiation produced at the radius $r=3$, corresponding to  the innermost stable circular orbit for a Schwarzschild BH.
In Fig.~\ref{fig:chis2} we show the angles $\chi^{\rm SR}$, $\chi^{\rm GR}$, and $\chi^{\rm tot}$ as a function of azimuth in the disc for different system inclinations. 
We define PA in the range $[-90\degr; 90\degr]$. 
In general, both the GR and SR rotations decrease with increasing inclination. 
This leads to higher depolarizing effects at smaller inclinations, which is a known result \citep{Dovciak2008}.

The curve $\chi^{\rm GR}$ (red dot-dashed in Fig.~\ref{fig:chis2}) is (anti-)symmetric with respect to the azimuth $\varphi=0\degr$ and $180\degr$. 
For $r=3$, the rotation is highest at $\varphi=180\degr\mp 17\fdg6$.  
The position of the extrema can be obtaied analytically if we apply \citet{B02} approximation for the light bending angle to the expression \eqref{eq:tanGR} for the rotation angle (see Appendix \ref{sec:app_gr} for derivation):  \begin{equation}\label{eq:cosphiGR_ext_main}
\cos\varphi^{\rm GR}_{\rm ext} = - \frac{r \sin^2 i +\cos^2 i}{r\sin i} .
\end{equation}
This approximation gives $\varphi^{\rm GR}_{\rm ext}=180\degr\mp 16\degr$. 


\begin{figure*}
\centering
\includegraphics[width=0.33\textwidth]{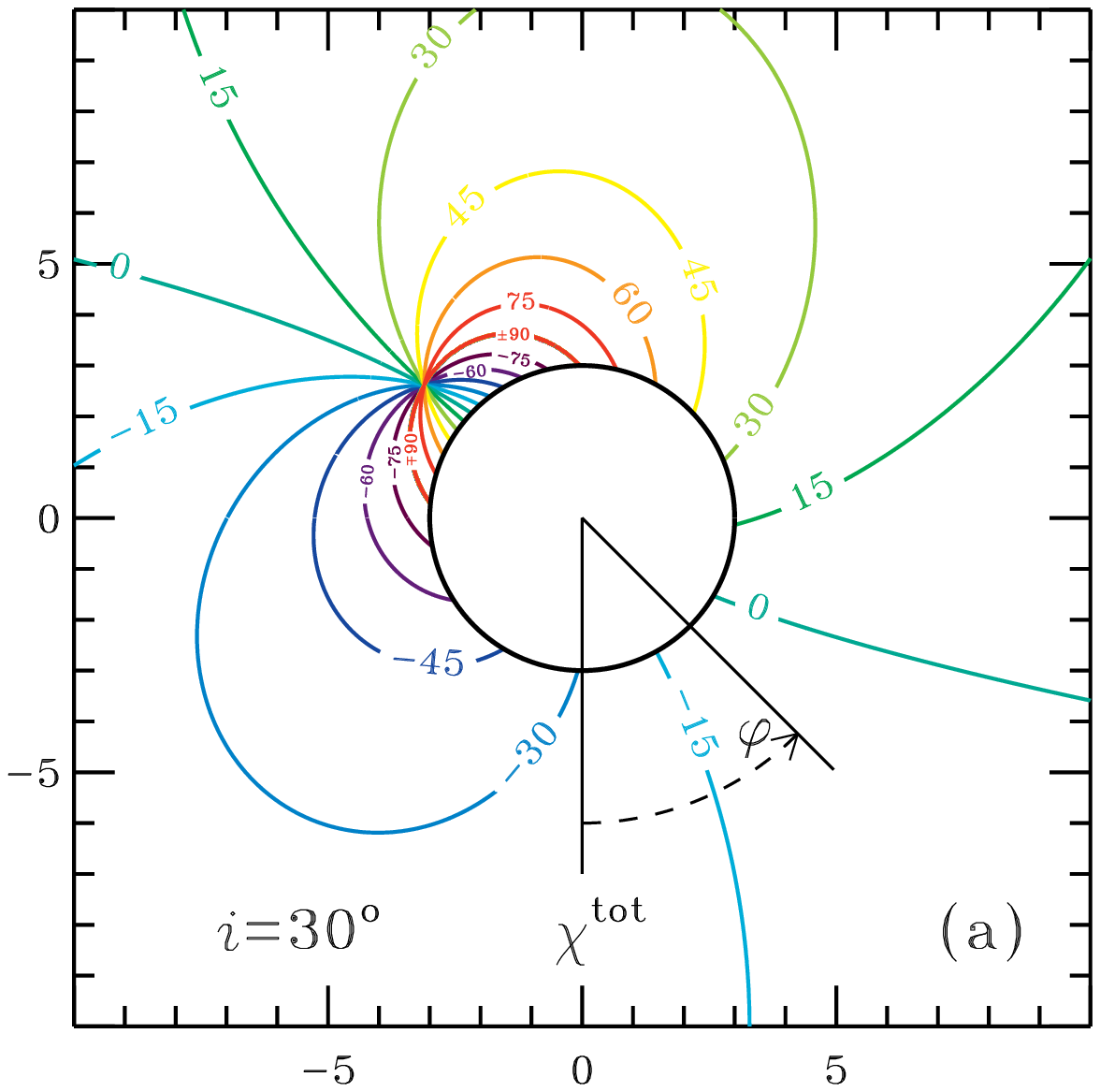}
\includegraphics[width=0.33\textwidth]{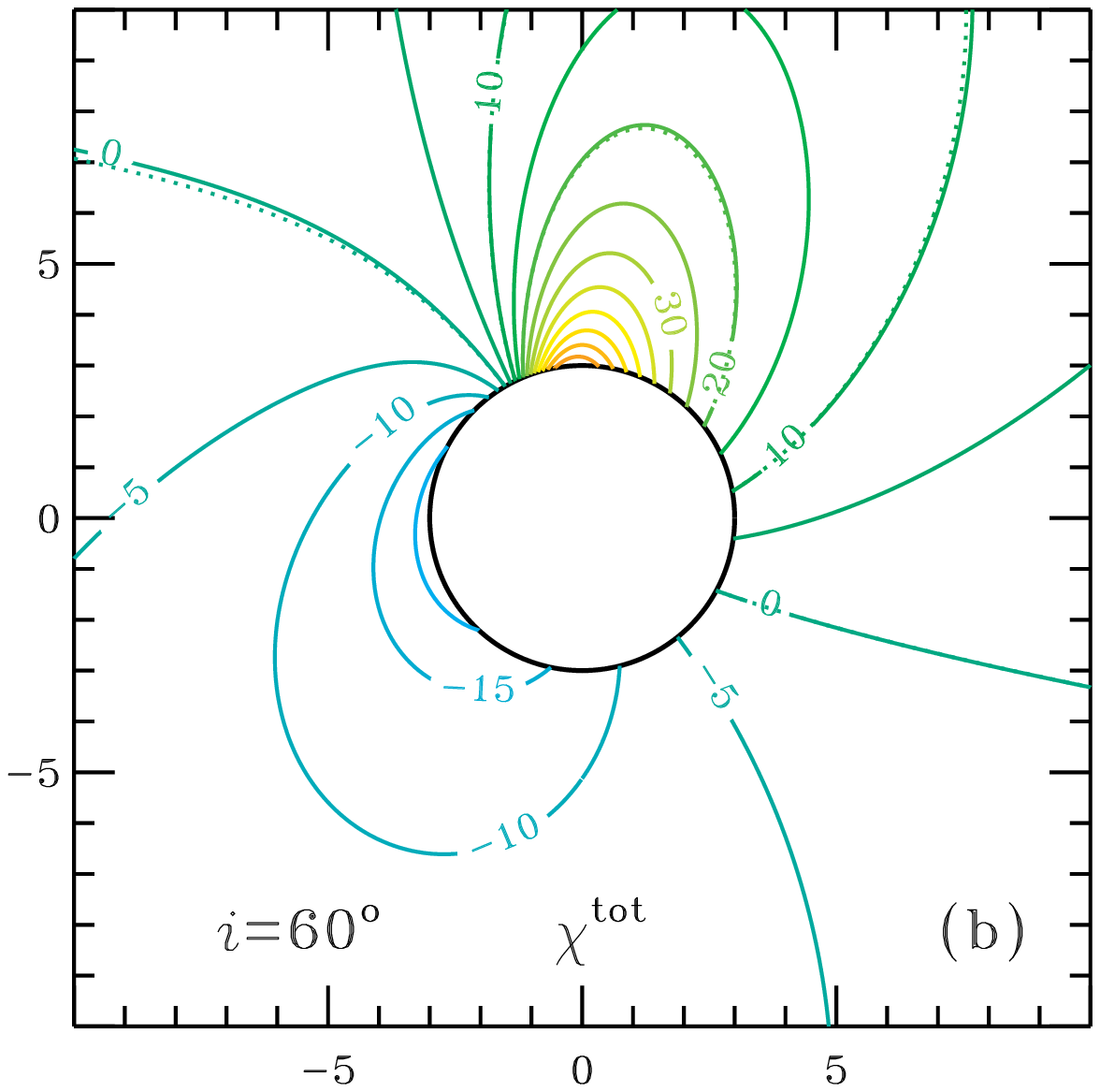}
\includegraphics[width=0.33\textwidth]{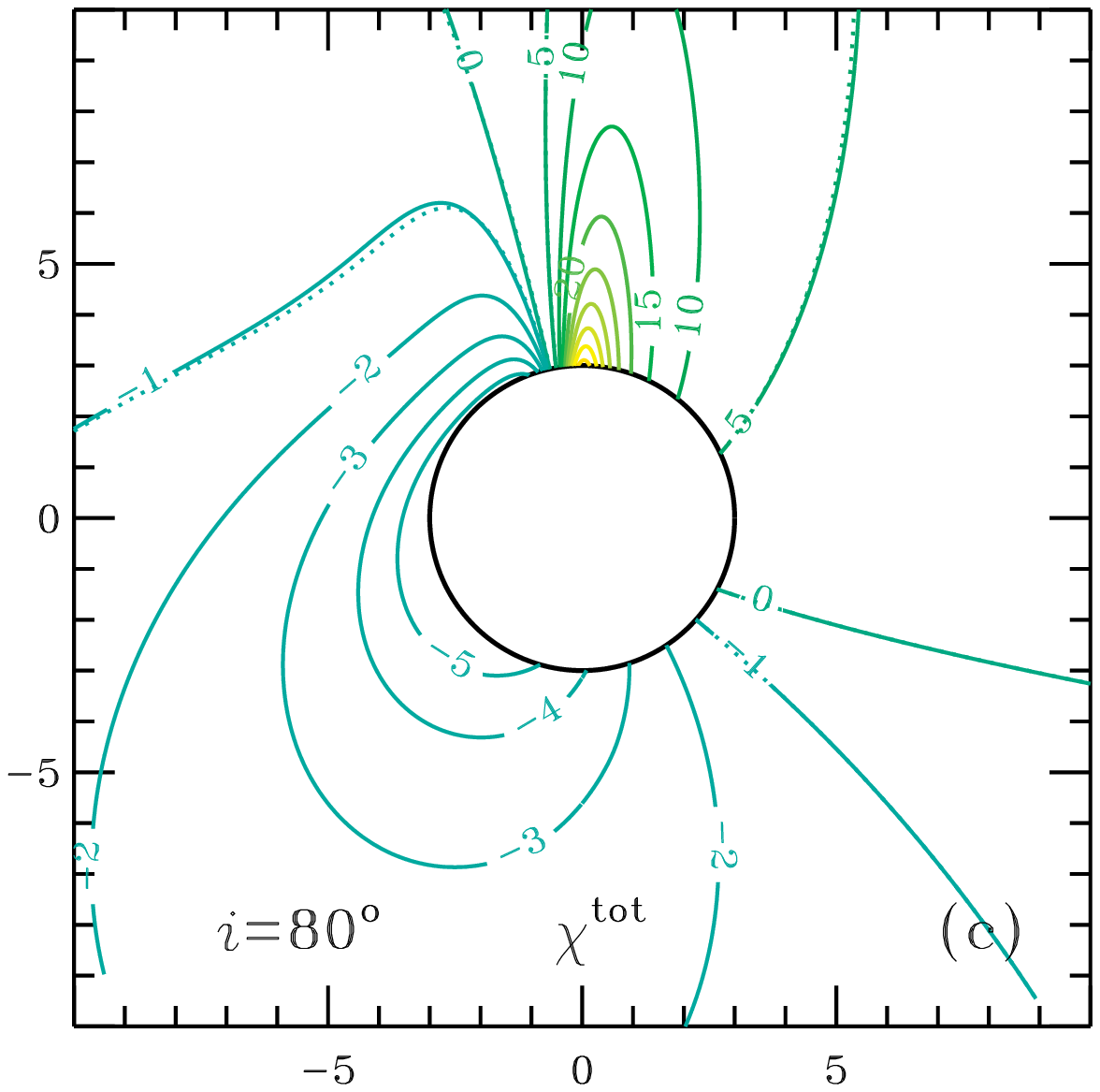}
\caption{
Contours of constant $\chi^{\rm tot}$ at the accretion disc plane $(r,\varphi)$. 
The observer is situated at inclinations $i=30\degr$ (\textit{panel a}), 60\degr\ (\textit{panel b}) and 80\degr\ (\textit{panel c}).
The coordinates are in units of $\rs$.
The polar angle $\varphi$ is measured from the projection of the direction to the observer on the disc plane. 
The disc rotates in the counterclockwise direction. 
The innermost stable circular orbit at $r=3$ is shown with black circle. 
The dotted lines (wherever visible) show corresponding contours computed using approximate formula for light bending \eqref{eq:bend_pout}.}
\label{fig:chitot_contours}
\end{figure*}
%

\begin{figure*}
\centering
\includegraphics[width=0.33\textwidth]{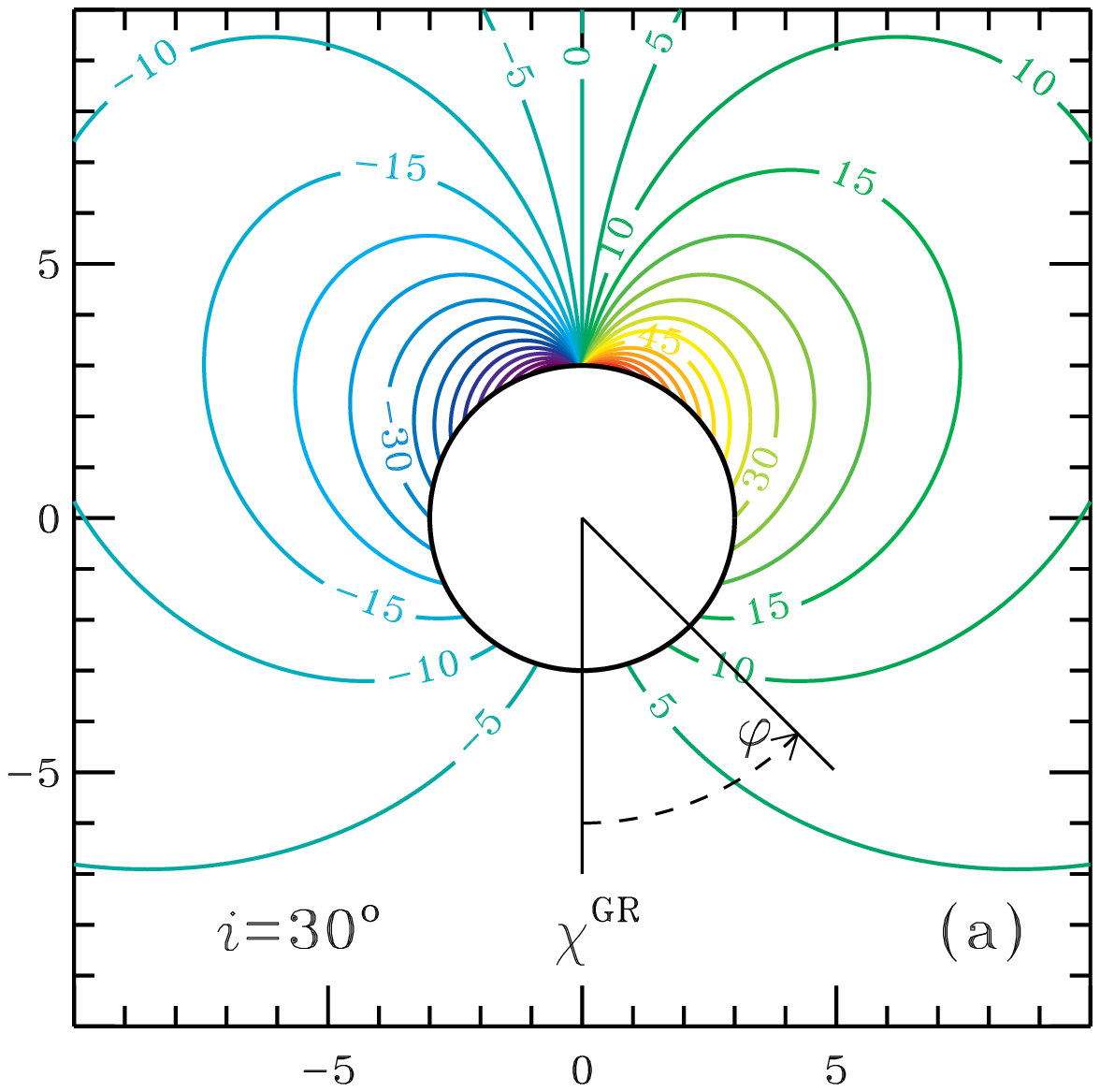}
\includegraphics[width=0.33\textwidth]{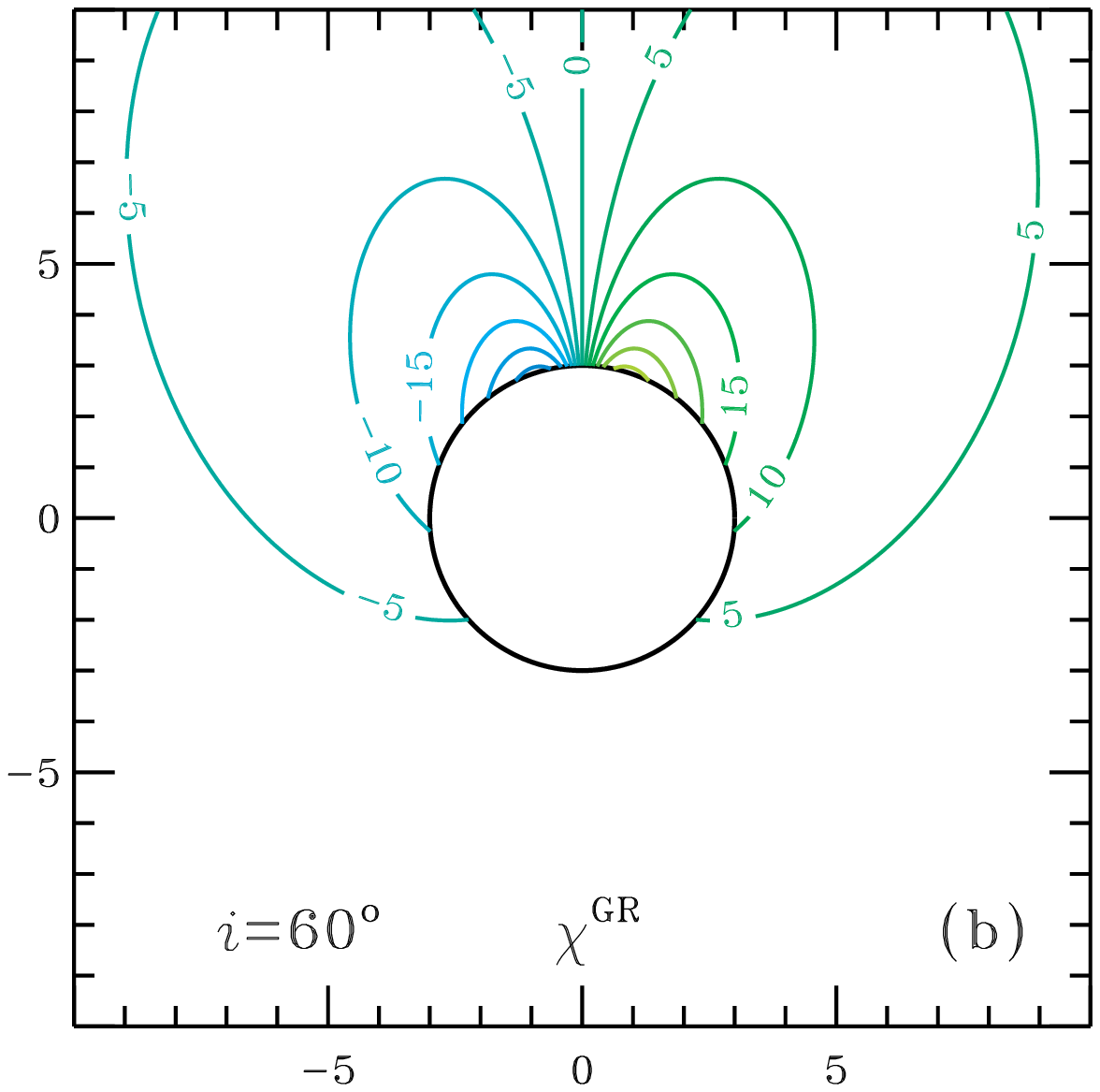}
\includegraphics[width=0.33\textwidth]{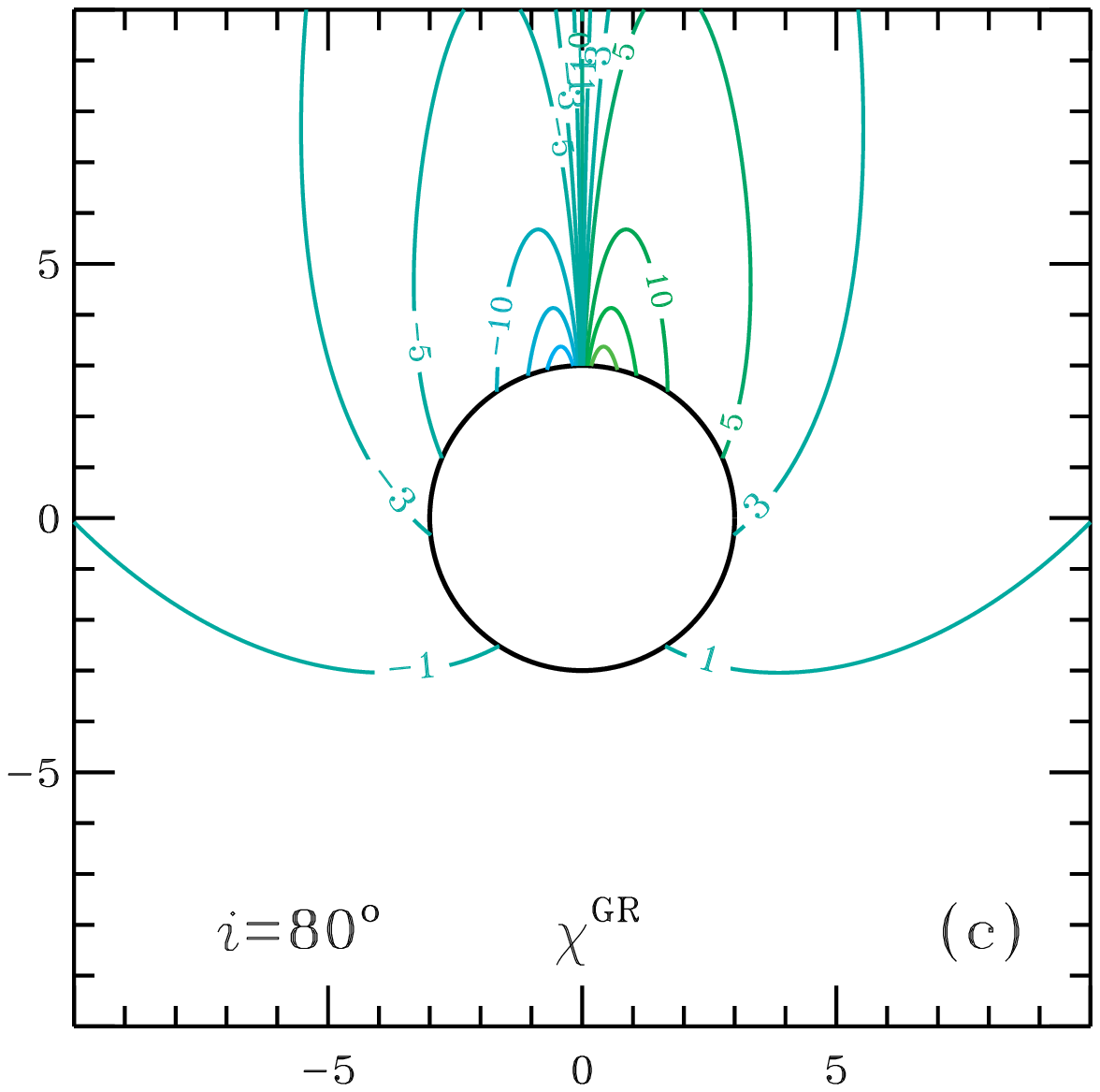}
\caption{Same as Fig.~\ref{fig:chitot_contours} but for $\chi_{\rm GR}$. }
\label{fig:chigr_contours}
\end{figure*}
%

\begin{figure*}
\centering
\includegraphics[width=0.33\textwidth]{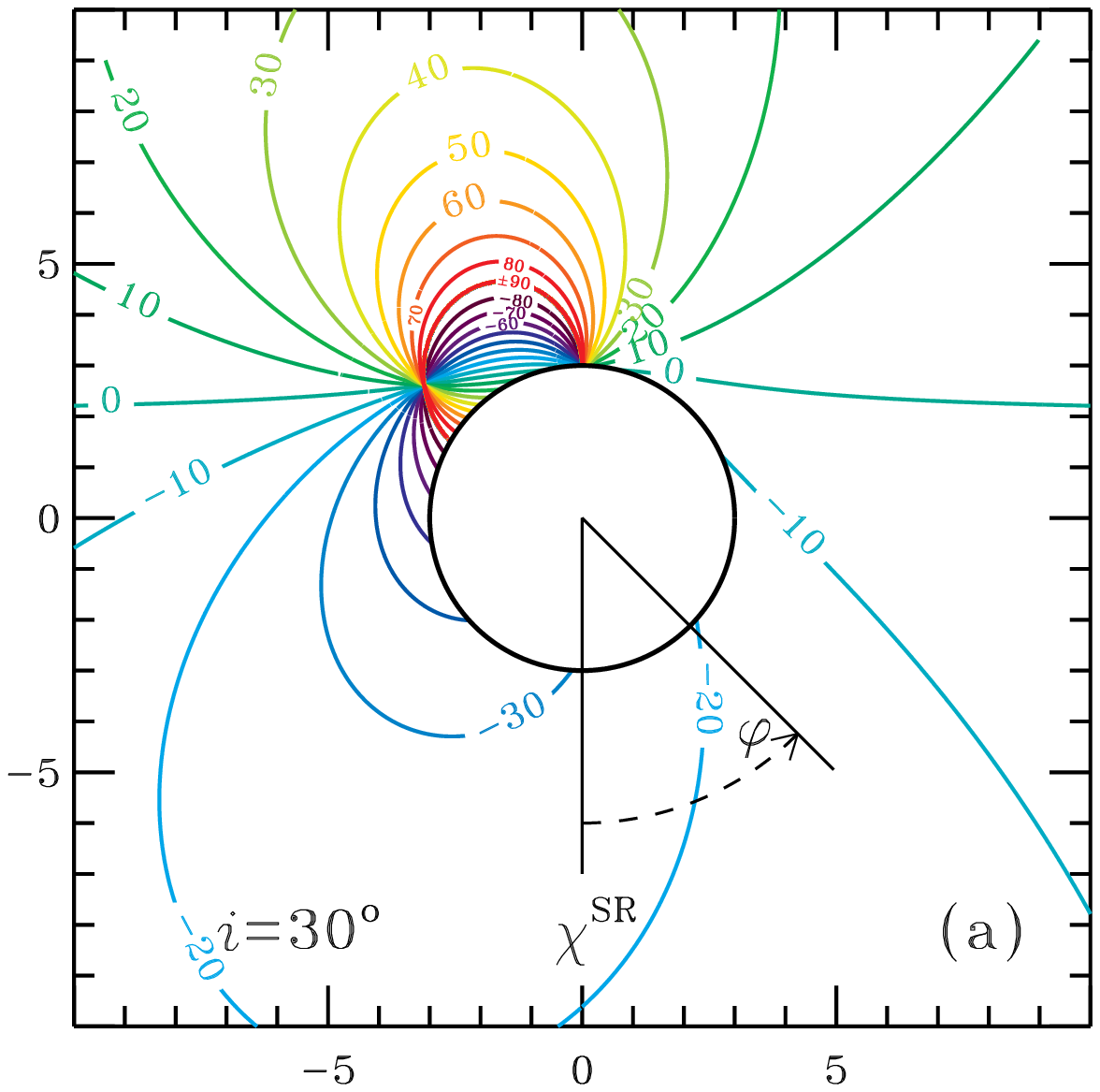}
\includegraphics[width=0.33\textwidth]{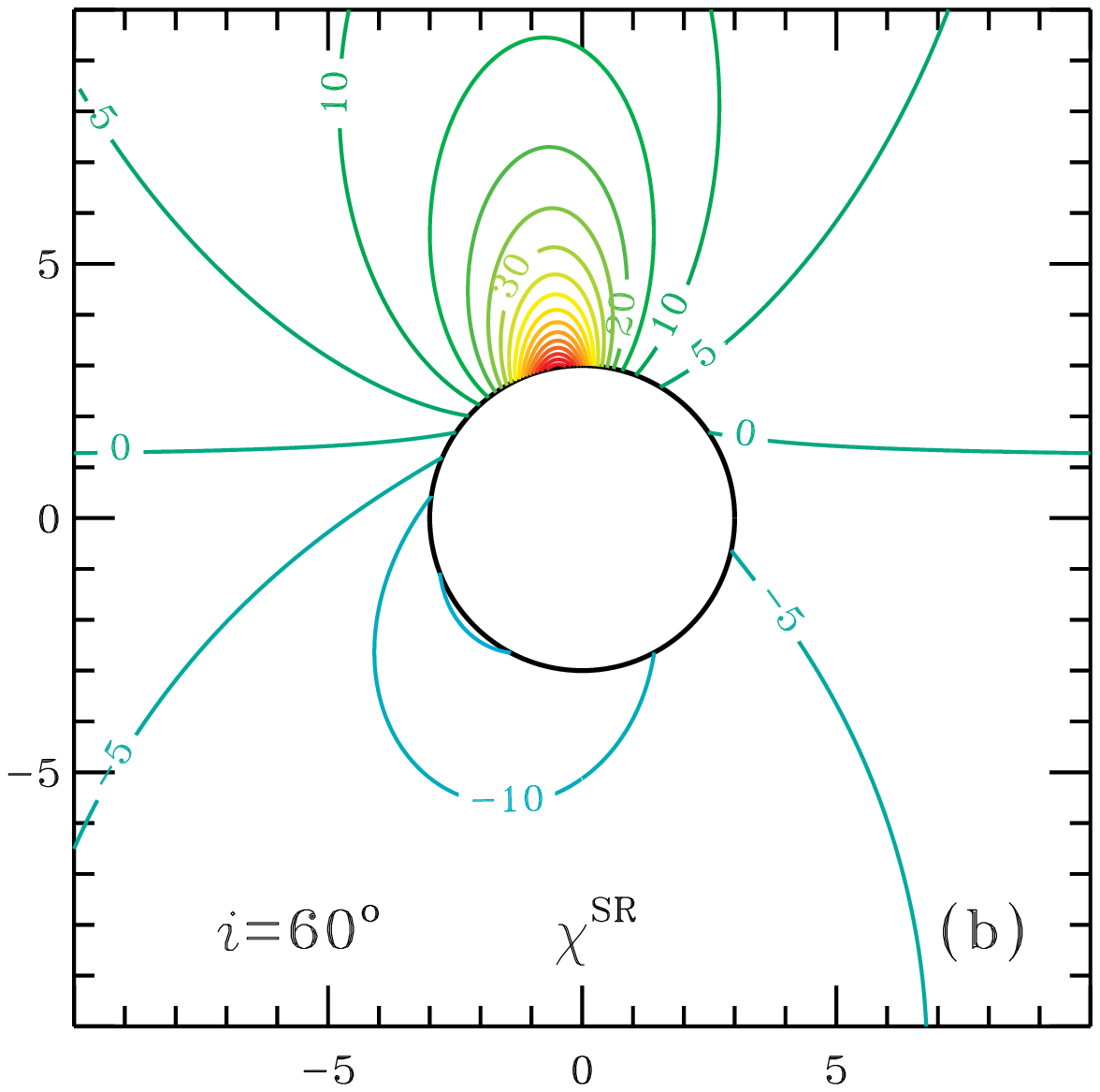}
\includegraphics[width=0.33\textwidth]{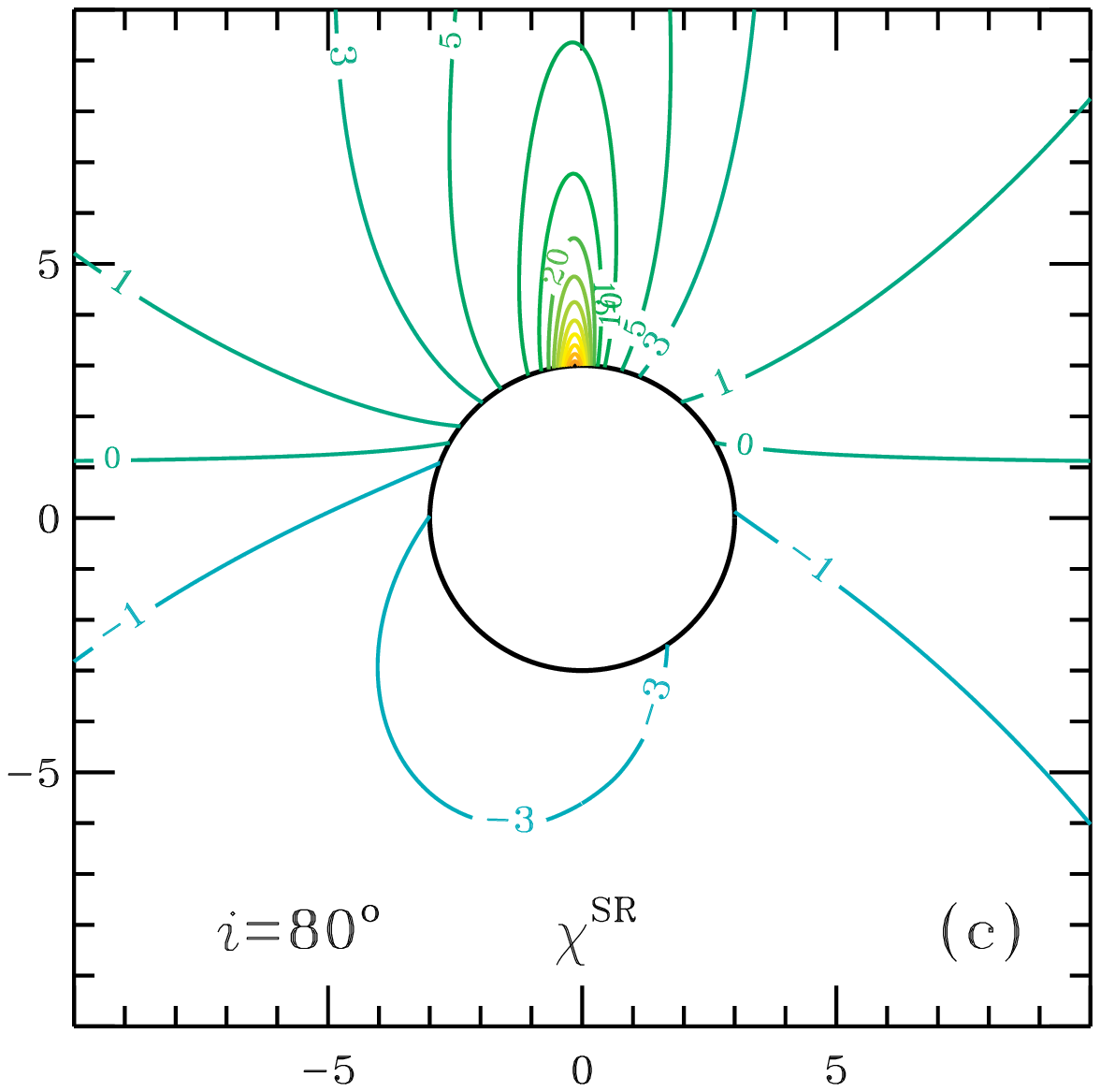}
\caption{Same as Fig.~\ref{fig:chitot_contours} but for $\chi_{\rm SR}$. }
\label{fig:chisr_contours}
\end{figure*}

The SR effects are computed for the Keplerian rotation given by Eq.~\eqref{eq:beta}. 
The curve $\chi^{\rm SR}$ (blue dashed curve in Fig.~\ref{fig:chis2}) is not symmetric, which adds to asymmetry of the total PA rotation $\chi^{\rm tot}$. 
At sufficiently large inclinations, $\chi^{\rm SR}$ has a maximum very close to $\varphi=180\degr$. 
Interestingly, the shape of the similar curve in flat space, $\chi^{\text{SR}}_{\rm flat}$, is very different (compare dashed blue with dotted green curves). 
In flat space, the maximum rotation angle is reached at (see Eq.~\ref{eq:tanSRflat}): 
\begin{equation}\label{eq:cosphiSR_flat_ext}
\sin\varphi^{\rm SR,flat}_{\max} = - \frac{\beta}{\sin i} .
\end{equation}
Obviously, the extrema do not exist if $\beta>\sin i$ and then $\chi^{\text{SR}}_{\rm flat}$ is monotonic. 
A general formula for the rotation angle $\chi^{\text{SR}}$ does not allow us to obtain a simple analytical expression for the position of the extrema. 
We note that the numerator of Eq.~\eqref{eq:tanSRGR} is a smooth function of $\varphi$, therefore, the rotation angle reaches maximum close to the azimuth where the denominator reaches minimum:  
\begin{equation}\label{eq:cosphiSR_ext}
\sin\varphi^{\rm SR}_{\max} \approx 
- \frac{\sin^2\zeta}{\beta\sin i\sin\alpha/\sin\psi} .
\end{equation} 
By noticing that the corresponding azimuth is close to $180\degr$ and that $\alpha$ and $\psi$ vary slowly with $\varphi$, and putting $\psi \approx 90\degr+i$, we get
\begin{equation}\label{eq:cosphiSR_ext2}
\sin\varphi^{\rm SR}_{\max} \approx 
- \frac{\cos^2\alpha\ \cos i}{\beta\sin i\ \sin\alpha} ,
\end{equation} 
where we can substitute $\cos\alpha= u-(1-u)\sin i$ using \citet{B02} approximation for the light bending angle. 
For $r=3$ and $i=60\degr$ ($80\degr$), the exact calculations give the maximum at $\varphi^{\rm SR}_{\max}=189\fdg7$ ($182\fdg7$), while Eq.~\eqref{eq:cosphiSR_ext2} gives 
$184\fdg1$ ($182\fdg2$). 

The total rotation $\chi^{\rm tot}$ is shown with the black solid lines in Fig.~\ref{fig:chis2}.
Notably, for inclinations $i\lesssim30\degr$, the PA rotation is monotonic with azimuth and crosses $90\degr$.
This means that the polarization direction significantly changes, making two full cycles as the azimuth varies from 0 to $360\degr$. 
Although in Fig.~\ref{fig:chis2}a we see that only $\chi^{\rm SR}$ makes two cycles, the total effect is not entirely caused by the relativistic rotation. 
Namely, for small inclinations, all the light rays coming to the observer were originally emitted along the direction of radius-vectors, i.e. $\unit{k}_{0}\cdot\unit{r}>0$, for any azimuth.
It means that the observer sees the ring from its outer parts, as if every disc element was located between the observer and the BH in the flat space.
This is taken into account in $\chi^{\rm SR}$, when we apply Lorentz transformation to these rays.
The direction of matter motion, with respect to the observer, varies between these disc segments, leading to variations of the angles between vectors $\unit{k}_{0}'$ and $\unit{k}_{0}$.
Hence, the SR effects only enhance the PA rotation caused by the light bending, so as if GR effects alone did not cause the full rotation of PA (such as in Fig.~\ref{fig:chis2}a), the additional rotation coming from relativistic motion results in the excess of the $90\degr$ boundary.
For the same reason we see two full rotations of PA for the case when the bright spot orbits the BH.
For inclination $180\degr-i$,  both the GR and SR rotation angles change sign. 

In Fig.~\ref{fig:chis} we show the total PA rotation $\chi^{\rm tot}$ for different radii and inclinations, as well as the error on the PA  computed using analytical expression for light bending \eqref{eq:bend_pout}, as compared to the exact, numerical solution.
We see that the relativistic effects decrease with radius, as expected. 
The largest rotation of the PA is observed at smaller inclinations. 
The difference between the PA computed using analytical formulae and numerically is always smaller than $\sim0\fdg6$.

The topology of $\chi^{\rm tot}$ is shown in Fig.~\ref{fig:chitot_contours} as a map on the accretion disc plane in the form of contours of constant values. 
It is similar to that shown in fig.~3 of \citet{Dovciak2008}, but defined here on the range of angles $[-90\degr,90\degr]$ (rather than the previously used $[-180\degr,180\degr]$, which included two full cycles of PA).
We see that for $i=30\degr$ there exists a critical point (at $r\approx 4.06, \varphi\approx 230\degr$) where  photons are emitted along the disc normal in the comoving frame (i.e. $\zeta'=0$) resulting in zero PD and not defined PA. 
At large inclinations, the strongest variations in PA are seen around $\varphi=180\degr$.
Calculations using analytical formula for light bending produce nearly identical picture (see the dotted lines barely separable from the solid lines in Fig.~\ref{fig:chitot_contours}).

\begin{figure*}
\centering
\includegraphics[width=0.95\textwidth]{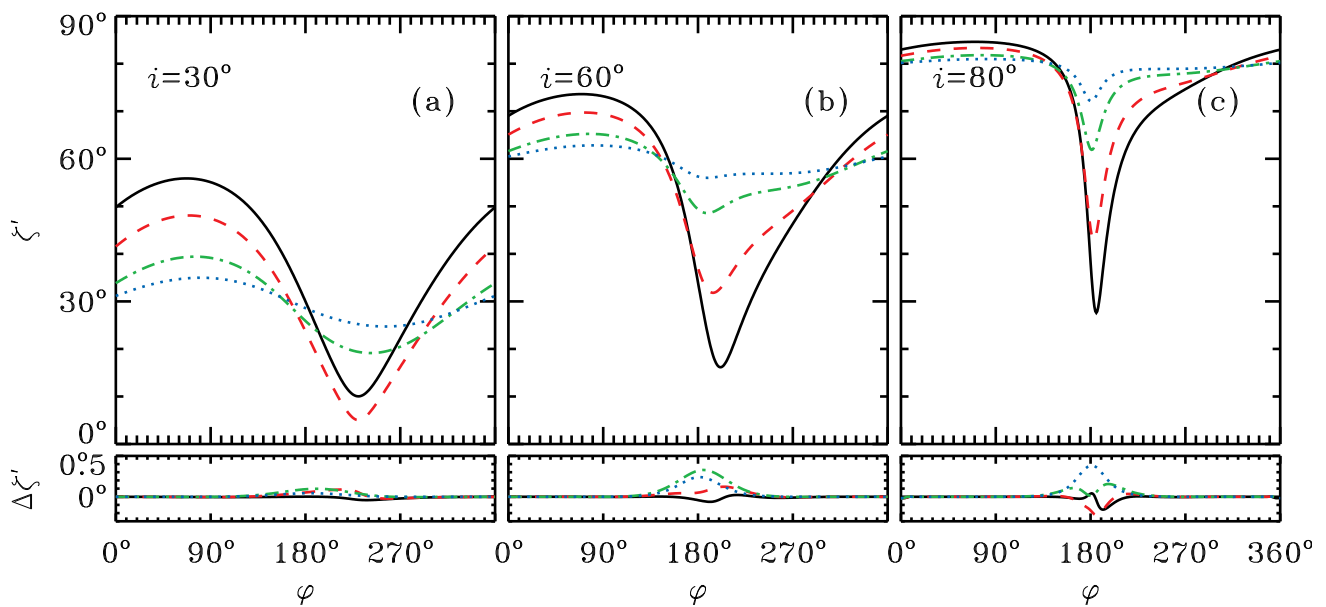}
\caption{\textit{Upper panels}:  Zenith angle $\zeta'$ the photon reaching the observer makes to the local normal in the fluid frame as a function of the azimuth for disc inclinations $i=30\degr$ (\textit{panel a}), $60\degr$ (\textit{b}) and $80\degr$ (\textit{c}) and different emission radii $r = 3$ (black solid line), 5 (red dashed), 15 (green dot-dashed) and 50 (blue dotted). \textit{Lower panels}:  Difference between $\zeta'$ computed using analytical formula \eqref{eq:bend_pout} for light bending and the one using exact formula. 
}
\label{fig:zetaprime}
\end{figure*}

The general topology of $\chi^{\rm tot}$ is easier to understand by looking separately at the topology of $\chi^{\rm GR}$ and $\chi^{\rm SR}$, which are presented in Figs.~\ref{fig:chigr_contours} and \ref{fig:chisr_contours}, respectively.  
The anti-symmetry of $\chi^{\rm GR}$ in $\varphi$ and very small rotation angles around $\varphi\approx 0$ are clearly seen. 
Also at $\varphi\approx 180\degr$ the GR rotation is small at large radii, while closer to $r=3$ the contours of constant values become highly packed resulting in a very fast changes of PA with azimuth.  
At small inclinations the GR rotation of PA is large, while, e.g. at $i=80\degr$ significant rotation happens only close to $\varphi=180\degr$. 
For the $\chi^{\rm SR}$, we see the existence of the critical point at small inclinations outside $r=3$. 
The peak in SR rotation is reached close to $\varphi=180\degr$ and the effect of SR rotation decreases with the increasing inclination, similarly to the GR rotation angle. 

\begin{figure*}
\centering
\includegraphics[width=0.95\textwidth]{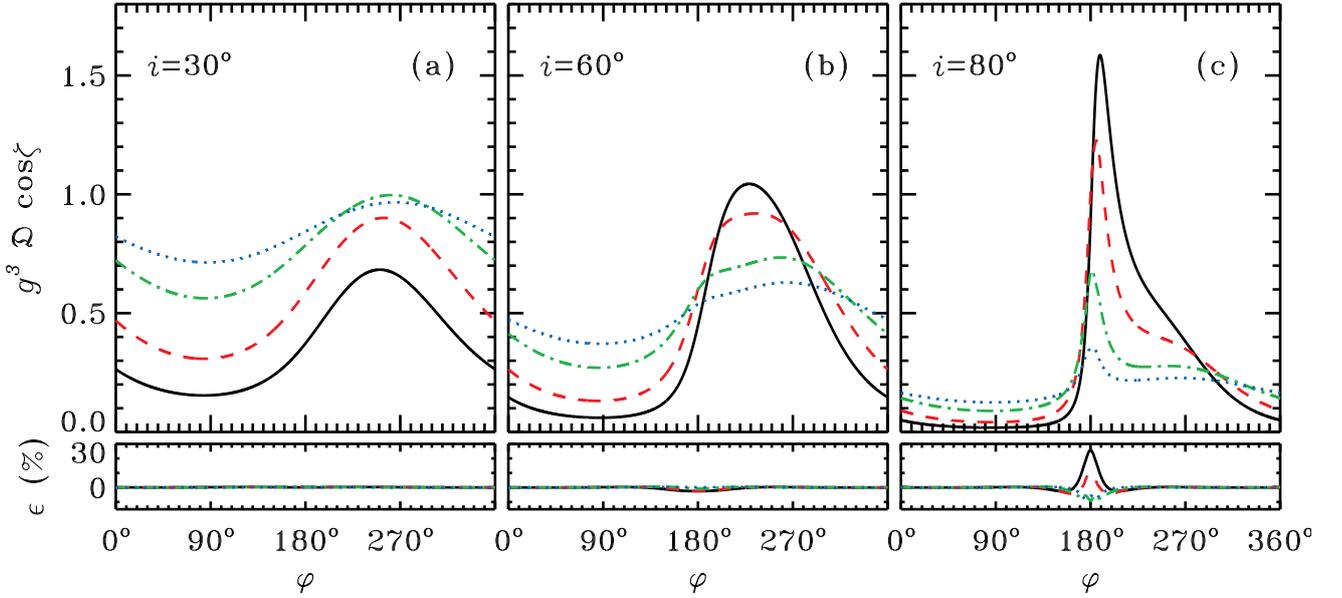}
\caption{ 
\textit{Upper panels}: Combination $g^3 {\cal D} \cos\zeta$ which is proportional to the observed flux from a disc surface element as a function of the azimuth for inclinations $i=30\degr$ (\textit{panel a}), $60\degr$ (\textit{b}) and $80\degr$ (\textit{c}) and different emission radii $r= 3$ (black solid line), 5 (red dashed), 15 (green dot-dashed) and 50 (blue dotted). 
\textit{Lower panels}: Relative error (in \%) on that combination computed using analytical formulae \eqref{eq:bend_pout} and \eqref{eq:lensing_pout} for light bending angle and lensing factor as compared to exact calculations.  
}
\label{fig:flux}
\end{figure*}

Interestingly, in the specific case of zero inclination $i=0$, the expressions for the GR and the SR rotation angles can be considerably simplified:
\begin{eqnarray}
\label{eq:chiGR_incl0}
\chi^{\rm GR} (r,i=0) & =&   \varphi, \\ 
\label{eq:chiSR_incl0}
\tan \chi^{\rm SR} (r,i=0) & =&  - \beta\ \tan \alpha, 
\end{eqnarray}
resulting in the total rotation 
\begin{equation}
\chi^{\rm tot} (r,i=0) = \varphi - \atan\left(\beta\,\tan \alpha\right).    
\end{equation}
Using \citet{B02} approximation and assuming Keplerian rotation, we further get: 
\begin{equation} \label{eq:chitot_izeroB02}
\chi^{\rm tot} (r,i=0) \approx \varphi  - \atan\left(\sqrt{\frac{1+u}{2u}}\right) = \varphi  - \atan\left( \sqrt{\frac{1+r}{2}} \right).    
\end{equation} 
For $r\rightarrow\infty$, this transforms to $\chi^{\rm SR} (r,i=0)=\varphi -\uppi/2$. 
This is different by $\uppi$ from the expression for the rotation angle in flat space given by Eq.~\eqref{eq:tanSRflat} in the limit $i\rightarrow 0$:
\begin{equation}
\chi^{\rm SR}_{\rm flat} (r,i=0) = \frac{\uppi}{2} + \varphi ,
\end{equation} 
i.e. both limits give the same position angle of the polarization (pseudo-)vector.
On the other end, at the innermost stable orbit, $r=3$, according to Eq.~\eqref{eq:chitot_izeroB02} the PA rotates by $\approx\varphi-54\fdg74$ (exact calculations give $\approx\varphi-54\fdg87$).

Light bending and aberration do not only rotate polarization plane, but also affect the flux and PD, as they alter the angle at which we see the surface element.
In Fig.~\ref{fig:zetaprime} we show, for the photon reaching the observer, the zenith angle $\zeta'$ between the local normal and photon propagation direction in the fluid frame, as well as the difference between the angles computed numerically and using approximate analytical formula \eqref{eq:bend_pout}.
This angle differs from the observer inclination $i$ because of the light bending and aberration.
The photon rays originating in the parts of disc behind the BH (around $\varphi = 180\degr$) experience the most pronounced bending, because the light trajectory lies above the BH at closer distances than the emission radius.  
This results in pronounced dips in $\zeta'$ seen in Fig.~\ref{fig:zetaprime}c, i.e. the disc here is seen more face-on (i.e. $\zeta'<i$).
The SR effects make the surface seen more face-on at places, where matter moves towards to the observer, that is $\varphi>180\degr$ and otherwise more edge-on for $\varphi<180\degr$, where the matter of the disc moves away from the observer. 
This effect is noticeable for moderate and low inclinations, thus, the change of $\zeta'$ in Fig.~\ref{fig:zetaprime}a is mainly due to SR effects. 
The error in $\zeta'$, when using the approximate bending formula, grows with inclination, but the largest difference is still below $\sim0\fdg4$ (for $i=80\degr$, see Fig.~\ref{fig:zetaprime}c).

There is a general decrease of the importance of GR and SR effects with increasing distance from the BH, hence the difference between the PAs computed using exact and approximate lensing formulae decreases with radius (Fig.~\ref{fig:chis}a).
However, counter-intuitively, we see in Fig.~\ref{fig:zetaprime}(b,c) that the error in $\zeta'$ grows with the radius of the ring around azimuthal angle $\varphi=180\degr$. 
For high disc inclination, the impact parameter of photons propagating towards the observer --- as well as the distance of closest approach to the central compact object --- are much smaller than the radius of the corresponding ring. 
This leads to a larger, although still within $\sim0\fdg5$, inaccuracy in $\zeta'$ for the approximation formula at high inclinations.

In  Fig.~\ref{fig:flux}, we show the quantity $g^3 {\cal D} \cos\zeta$, which is proportional to the observed flux from a surface element  (see Eq.~\ref{eq:fluxspot}). 
Here, we clearly see the effect of Doppler boosting for part of the ring at  $\varphi>180\degr$ and deboosting for $\varphi<180\degr$. 
For high inclination observer, gravitational lensing strongly amplifies the flux from the part of the disc behind the BH at $\varphi\sim 180\degr$. 
The azimuthal dependence of the flux is not symmetric because of the increasing role of the Doppler boosting towards $\varphi\sim 270\degr$. The relative error on the flux arising from the approximate bending and lensing formulae is largest for high inclinations, at azimuths $\varphi\sim 180\degr$, when the angle $\psi$ is large. 
The largest error at $r=3$ reaches 26\% for $i=80\degr$, while for $i=30\degr$ and $60\degr$ it does not exceed 0.6\%.

\begin{figure}
\centering
\includegraphics[width=0.9\linewidth]{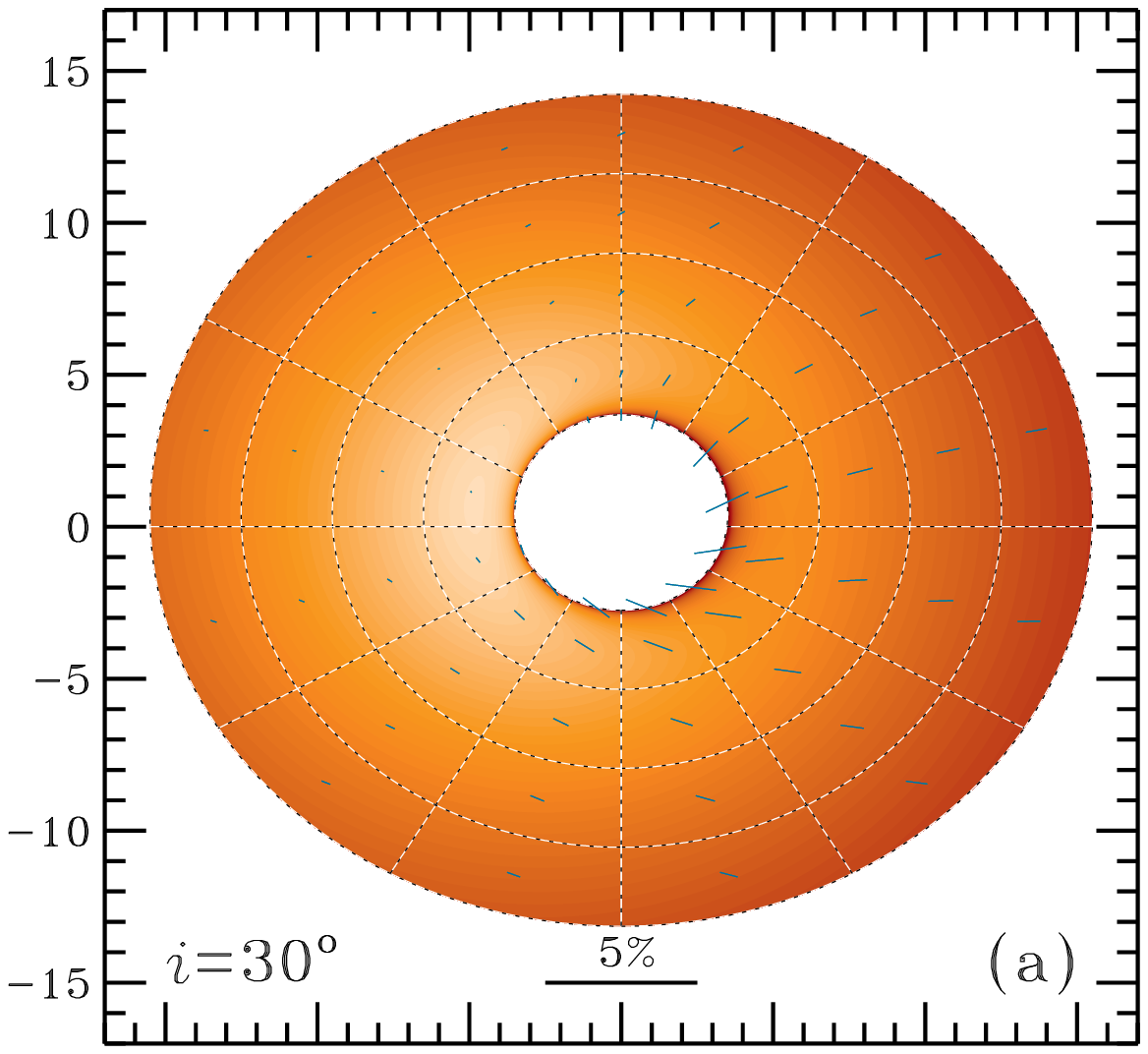}
\includegraphics[width=0.9\linewidth]{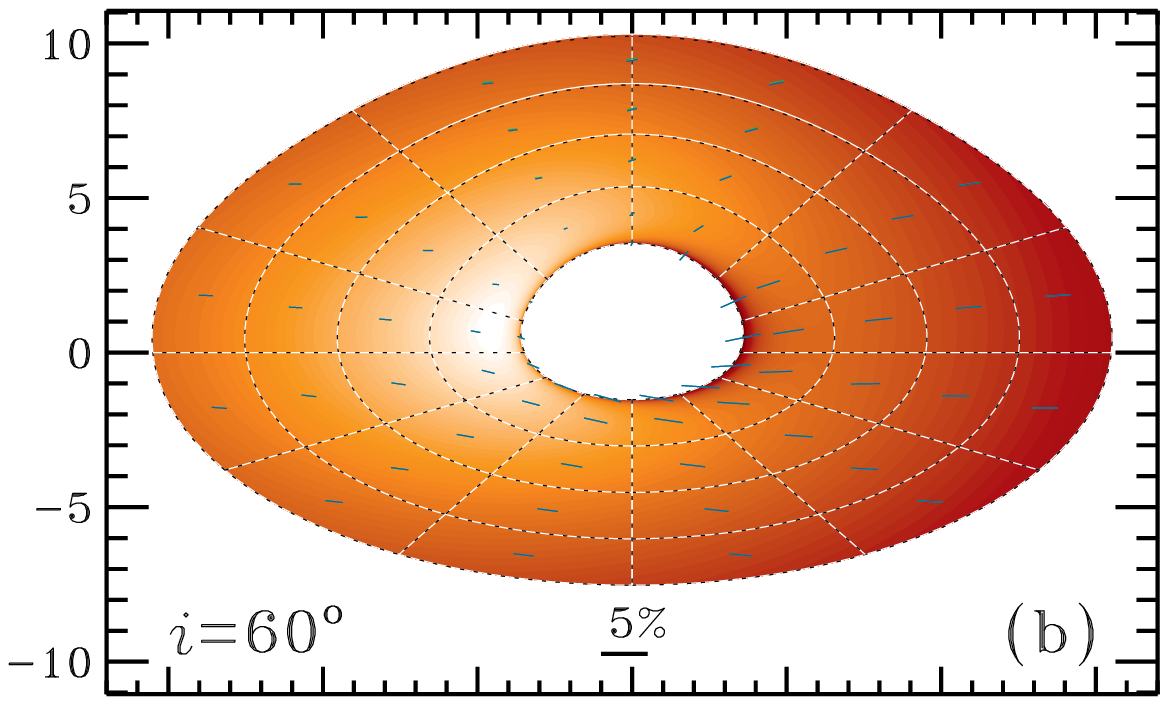}
\includegraphics[width=0.9\linewidth]{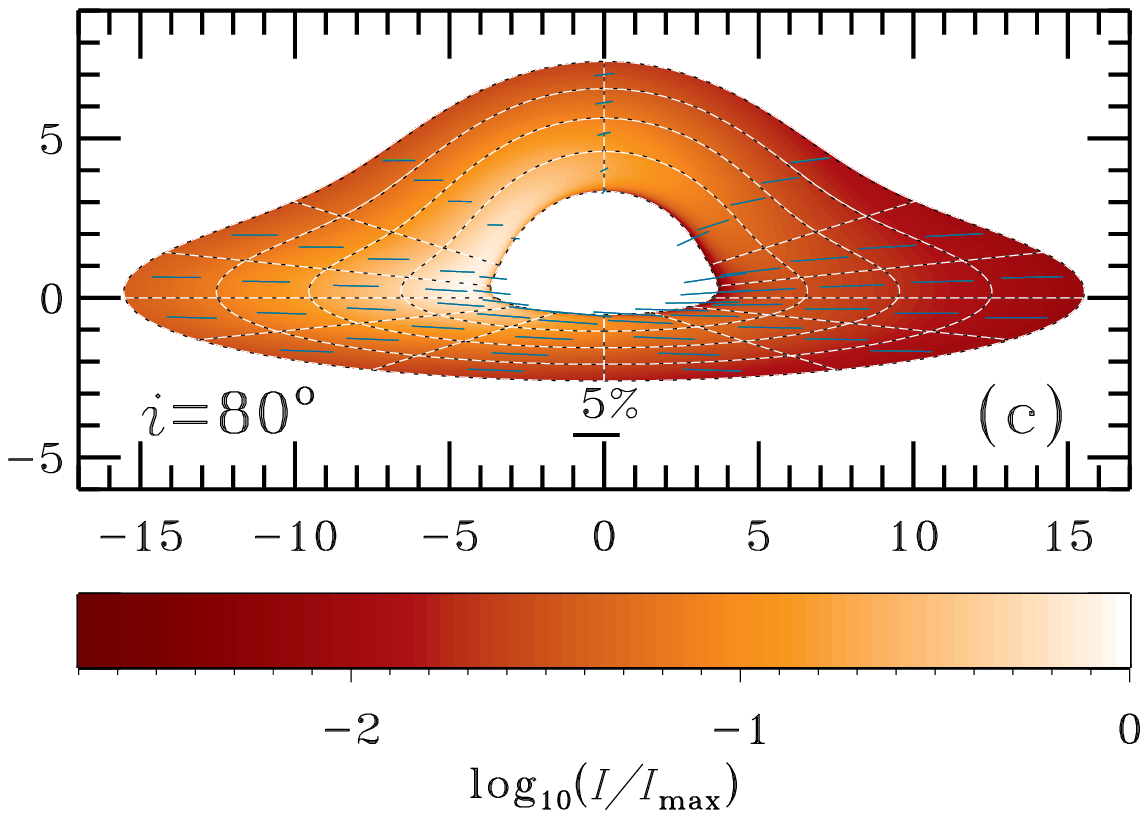}
\caption{Images of an accretion disc as viewed at three inclinations $i=30\degr$ (\textit{panel a}), 60\degr\ (\textit{b}) and 80\degr\ (\textit{c}).  
The coordinates on the sky are given in units of $\rs$. 
Contours show the images of rings of equal radii $r$ (from 3 to 15, with step of 3) and equal azimuths $\varphi$ (every 30\degr). 
Black solid contours correspond to the exact calculations of $\alpha(\psi)$, while the white dashed contours (almost fully overlapping with the black ones) are for the approximation \eqref{eq:bend_pout}. 
The colours reflect the logarithm of $g^4 t^4(r) a_{\rm es}(\zeta')$, which is proportional to the bolometric intensity. 
The blue sticks give polarization computed using approximate light bending expression \eqref{eq:bend_pout}, with their length being proportional to the PD and their position angle is given by the PA. 
Exact calculations shown by green sticks are nearly indistinguishable from the approximation.  }
 \label{fig:disc}
 \end{figure}

\subsection{Polarization of accretion disc in Schwarzschild metric}

We compute the polarization signatures of the optically thick, geometrically (infinitely) thin accretion disc using the analytical formulae derived above.
The intensity of the disc in the fluid frame depends on the radius $r$, energy $E'$ and the angle $\zeta'$ between the photon vector and the disc normal, $\cos\zeta'=\unit{k}_{0}'\cdot \unit{n}$.
As an illustration, we consider the simple case of the standard accretion disc in Newtonian gravity \citep{SS73} and pure electron scattering atmosphere for polarization properties.
Substituting relevant dependencies in Eq.~\eqref{eq:primepolvec}, we obtain the Stokes vector in the fluid frame in the form
\begin{equation} \label{eq:primepolvec2}
  \bm{I}'_{E'}(r,\zeta')=\frac{1}{f_{\rm c}^4} B_{E'}\left(T_{\rm c}(r)\right) a_{\rm es}(\zeta')
  \begin{bmatrix} 1\\ - p_{\rm es}(\zeta')\\0  \end{bmatrix},
\end{equation}
where 
\begin{equation}
  a_{\rm es}(\zeta') \approx 0.421+0.868\,\cos\zeta'
\end{equation}
approximates the angular dependence of the outgoing intensity \citep{SPW20}, 
\begin{equation}
    p_{\rm es}(\zeta') \approx 11.71\% \frac{1-\cos\zeta'}{1+3.5 \cos\zeta'} 
\end{equation}
approximates the angular dependence of the PD, and the PA of $\chi_0=\uppi/2$ is used corresponding to the direction of polarization vector perpendicular to the meridional plane \citep{Chandrasekhar1947,Cha60,sob49,Sob63}.

The spectral shape is described by the Planck function $B_{E'}$ of the colour temperature $T_{\rm c}=f_{\rm c}T_{\rm eff}$  with the colour correction factor assumed to be $f_{\rm c}=1.7$ \citep{ST95}.
For the radial dependence of the effective temperature we use a simple expression for a standard Newtonian disc \citep{SS73}:  
\begin{equation}\label{eq:tempprofile}
    T_{\rm eff}^4(r) = \frac{3GM\dot{M}}{8\pi \sigma_{\rm SB} R^3} \left(1-\sqrt{\frac{3\rs}{R}}\right) = T_*^4 t^4(u), 
\end{equation}  
where 
\begin{equation}\label{eq:tempstar}
    T_*^4  = \frac{3GM\dot{M}}{8\pi \sigma_{\rm SB} \rs^3} ,
\end{equation} 
\begin{equation}\label{eq:tu_ss73}
  t (u) = \left[u^3 \left(1-\sqrt{3u}\right)\right]^{1/4} .
\end{equation} 

Fig.~\ref{fig:disc} shows images of the accretion disc as viewed at three different inclinations. 
The colours reflect the bolometric intensity, which is given by the product $g^4 t^4(r) a_{\rm es}(\zeta')$. 
The black and white contours represent the sky images of lines of equal radii $r$ and equal azimuths $\varphi$, computed using exact bending relation and its approximation \eqref{eq:bend_pout}, respectively. 
The difference is only visible for $i\geq60\degr$ at the accretion disc side behind the black hole, $\varphi\approx 180\degr$, which correspond to $\psi\approx$150\degr--170\degr\ and photon trajectories lying close to the black hole. 
We also plot there polarization pseudo-vectors.
The green sticks correspond to the exact calculations, while the blue one to the approximate light bending formula. 
The PA and PD  are nearly identical for these two cases, as they are defined by $\chi^{\rm tot}$ and $\zeta'$, which are very well approximated by the analytical formulae (see  Figs.~\ref{fig:chitot_contours} and \ref{fig:zetaprime}). 
Small inaccuracy of the light bending formula leads to minor difference in the positions of the sticks on the plane of the sky, while the difference in their angles (PA) and lengths (PD) cannot be seen by eye. 

Using Eqs.~\eqref{eq:polflux} and \eqref{eq:Stokes_spot2}, we integrate over radius and azimuth to get the observed Stokes vector.  
The Stokes vector of the apparent luminosity is computed by multiplying the Stokes vector $\bm{F}_E$ by $4\pi D^2$.
The luminosity can be represented in a dimensionless form by scaling it to $\sigma_{\rm SB}T_*^4\rs^2$, multiplying by photon energy and using the dimensionless photon energy $x=E/kT_*$ as an argument: 
\begin{eqnarray}
 \label{eq:lum_scaled}
x \bm{l}_x & = &  \frac{x \bm{L}_x}{\sigma_{\rm SB}T_*^4\rs^2} = \frac{60}{\pi^4} 
 \int\limits_{r_\textrm{in}}^{r_\textrm{out}}\!\!   \frac{r \,\rmd r}{\sqrt{1-u}} 
 \int_0^{2\uppi} \rmd \varphi  \, 
\frac{g^4}{f_{\rm c}^4}\ {\cal D}\ \cos\zeta  \nonumber \\
& \times &
\frac{x'^4 a_{\rm es}(\zeta')}{{\rm e}^{\displaystyle x'/f_{\rm c}t(u)}-1} \begin{bmatrix} 1\\ 
p_{\rm es}(\zeta') \cos(2[\chi^{\rm tot}+\chi_0]) \\ 
p_{\rm es}(\zeta') \sin(2[\chi^{\rm tot}+\chi_0]) \end{bmatrix} ,
\end{eqnarray}  
where $x'=gx$.  
With such scaling of photon energy and luminosity, spectra of accretion disc become independent of the BH mass and accretion rate. 

The positively-defined observed PD is then 
\begin{equation}\label{eq:pol_deg_total}
 p(x)= \frac{\sqrt{l_{x,Q}^2+l_{x,U}^2}}{l_{x,I}}  . 
\end{equation}
The observed PA is computed as an argument of the complex quantity formed by the  Stokes parameters: 
\begin{equation}\label{eq:pol_ang_total}
 \chi(x)= \frac{1}{2}\arg\ (l_{x,Q} + {\rm i} l_{x,U}) .
\end{equation}

\begin{figure}
\centering
\includegraphics[width=0.95\linewidth]{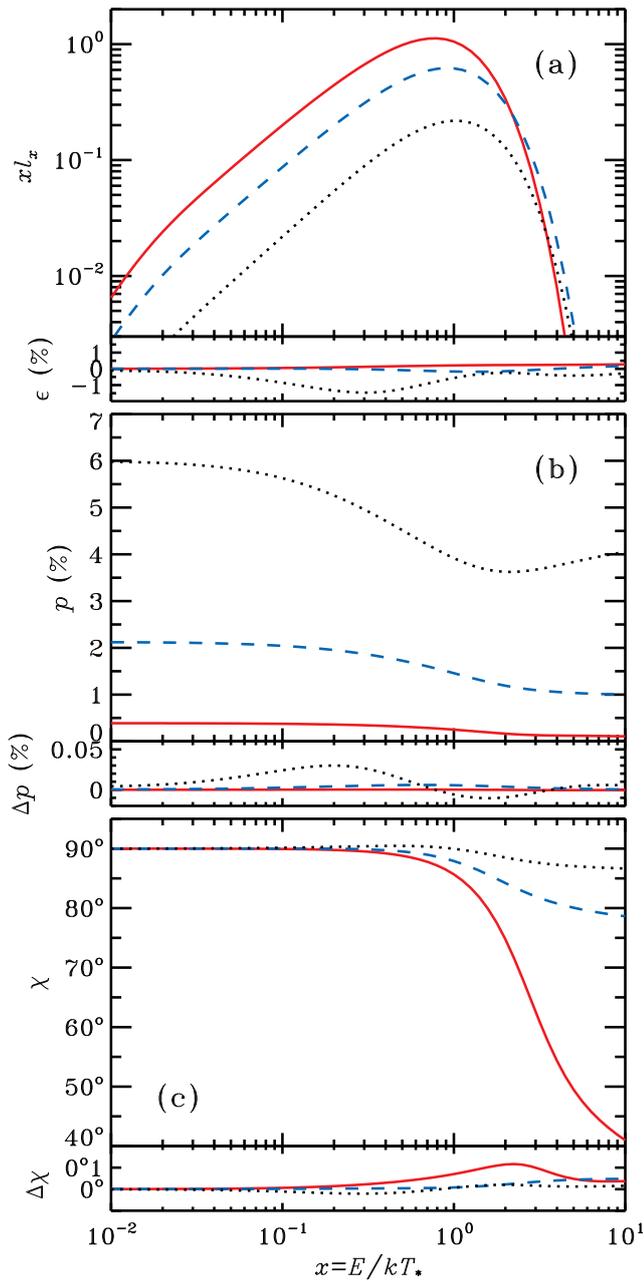}
\caption{Polarized spectrum of a standard accretion disc ($3<r<3000$) viewed at different inclinations: $i=30\degr$ (red solid line), $60\degr$ (blue dashed), $80\degr$ (black dotted) as function of dimensionless photon energy $x=E/kT_*$. 
\textit{Panel a:} Normalized luminosity $xl_x$. 
\textit{Panel b:} Polarization degree. 
\textit{Panel c:} Polarization angle. 
The lower subpanels show the errors in the corresponding quantities when computations are done with approximate expressions Eqs.~\eqref{eq:bend_pout} and \eqref{eq:lensing_pout} for the bending angle and lensing factor.   
We note that the error on $xl_x$ is relative, while the errors on PA and PD are absolute. }
\label{fig:stokes}
\end{figure}

In Fig.~\ref{fig:stokes} we show the resulting spectral energy distributions of $x\,l_x$, PD ($p$) and PA ($\chi$) for the accretion disc extending from $r_{\rm in} = 3$ to $r_{\rm out}=3000$ in Schwarzschild metric, as seen by a distant observer at different inclinations: $i=30\degr$, $60\degr$ and $80\degr$. 
Fig.~\ref{fig:stokes}a shows the dimensionless luminosity as a function of photon energy at three inclinations. 
It is much higher at low inclinations due to the strong beaming of radiation along the normal in the electron-scattering dominated atmosphere. 
At higher inclinations, the Doppler effects shifts the peak of emission to higher energies.   
The lower subpanel shows the relative error on luminosity when computations are done using approximate formulae for the light bending angle \eqref{eq:bend_pout} and lensing factor \eqref{eq:lensing_pout}. 
In spite of the fact that the error on the lensing factor reaches nearly 30\% at $r=3$ for $i=80\degr$ (see Fig.~\ref{fig:flux}), the integral flux has at most 1.5\% relative error, owing to the fact that the error in ${\cal D}$ at different radii has different sign.  

The PD (see Fig.~\ref{fig:stokes}b) is much higher for high inclinations. 
At lower energies, emission is dominated by large radii where relativistic effects are not important and PD follows the angular dependence given by $p(i)\approx p_{\rm es}(\zeta')$. 
In agreement with previous findings  \citep{StarkConnors1977,Connors1980,Dovciak2008}, we observe the increase of value of the rotation angle and increasing role of depolarization effects at higher energies (see Fig.~\ref{fig:stokes}b,c), where the emission from the innermost radii is most important. 
Because the observed flux is dominated (due to the Doppler effect) by the part of the disc at $\varphi\approx 270\degr$, where the rotation angle is large, the integrated rotation angle is high at small inclinations. 
The range of rotation angles is wide in this case  (see Figs.~\ref{fig:chis2}-\ref{fig:chitot_contours}), hence the depolarization effect is strongest at higher energies. 
As described, for example, in \citet{Dovciak2008}, the recovery of PD at energies higher than $kT_*$ is caused by the fact that only a small area of the disc contributes to this range of energies, therefore the depolarizing effects are smaller.
Calculations using approximate formulae for light bending and lensing factor give very accurate results in these cases, too. 
For example, the absolute error on PD is smaller than 0.03\% for all inclinations and the error on the PA barely exceeds $0\fdg1$.

\subsection{Extensions of the model} 

The fully analytical formalism developed above can naturally be applied to a number of problems on production of polarized radiation near relativistic objects.
First, the assumption that radiation is produced in a  plane-parallel electron-scattering atmosphere can be relaxed.
In this specific model, the Stokes vector defined in a polarization basis connected to the local normal contains only two non-zero parameters: $I$ and $Q$. 
Properties of the local emission can be very different in other setups. 
For example, the accretion flows around BHs at low accretion rates are optically thin and polarization at low photon energies (radio to sub-millimeter for super-massive BHs and optical--infrared for X-ray binaries) may be related to synchrotron radiation of relativistic electrons \citep{PV14,YN14}. 
The magnetic field direction in this situation will not likely be aligned with the local normal. 
In this case, the Stokes vector will contain three (or four, if we also consider circular polarization) components. 
In the X-ray domain, the PD and PA may also be very different from the optically thick case, as the photons are produced by Comptonization in a geometrically thick, optically thin hot flow.  
However, these deviations will change only the Stokes vector of the radiation escaping the disc, but will not affect the computational scheme of the observed polarized flux Stokes vector, which still  will follow Eq.~\eqref{eq:polflux}, with the expressions for the PA rotation angle remaining the same. 

Another extension of the model is related to the velocity field of the accreting matter. 
In this paper, we assumed that the gas velocity has only azimuthal component, while, for example, an optically thin hot  flow at low accretion rates, as seen by the EHT in M87, may have a significant radial component \citep{EHT21a,EHT21b,Narayan21}.
We note that rotation of the polarization plane due to light bending $\chi^{\text{GR}}$ does not depend at all on the velocity field. 
To calculate the rotation $\chi^{\text{SR}}$ caused by special relativity effects in this case, we can use a more general expression \eqref{eq:tanSRgeneral} in the vector form. 

Further extension of the model concerns the properties of the polarized flux if the emission is confined to small region (hot spots) corotating with the flow \citep[e.g.][]{Pineault1977pol_schw,Connors1980}. 
In this case, Eq.~\eqref{eq:Stokes_spot} for the observed Stokes vector as a function of emission azimuthal angle $\varphi$ should be modified to account for the light travel time delays. 
It will depend on the photon arrival time (corresponding to the phase of the orbit, $\varphi_{\rm obs}$):
\be \label{eq:Stokes_spot_lc}
\df{\bm{F}}_{E} (r,\varphi_{\rm obs}) =  g^{3}\ {\bf M}(r,\varphi)\ \bm{I}'_{E'}(\zeta')\ 
 \frac{\df{S'} \cos\zeta'}{D^2}\ {\cal D}, 
\ee
where $\df{S'}$ is the emission region area in the fluid frame, which is now multiplied by $\cos\zeta'$, instead of $\cos\zeta$. 
The relation between the emission azimuth $\varphi$ and the arrival time can be obtained from the time delay integral \citep[see e.g.][]{PFC83,PB06,SNP18}.

\section{Summary}

In this work we derived explicit analytical expressions describing the rotation of the polarization plane in Schwarzschild metric.
We showed that the total rotation is a sum of two effects: the rotation caused by the light bending (Eq.~\ref{eq:tanGR}), which is a pure GR effect, and the relativistic aberration, the latter has to be applied to the photons whose path will subsequently be altered by the bending (Eq.~\ref{eq:tanSRGR}).
The latter effect acts on the PA in a way that is different from the relativistic aberration in flat space.

We show test cases for the observed  PA from a disc ring, as a function of azimuth for various emission radii and observer inclinations.
Combining with the recently derived analytical formula for the light bending, it is possible to compute both the zenith angle of emission in the fluid frame (that will affect the observed flux and PD) and PA with an accuracy better than 1\degr\ for any inclination and emission radius. 
This {opens a possibility to}   
produce high-precision polarized images of the accretion discs in Schwarzschild metric in unprecedentedly small computing time. 

Integration over the disc surface gives the full observed Stokes vector as a function of energy. 
Utilizing the analytical formulae allows to reduce the computing time compared to the ray-tracing calculations by a factor of a hundred giving accuracy better than 1\% on the flux, 0.03\% on the integrated PD and about $0\fdg1$ for the PA. 
Our analytical technique can be used for detailed comparison of theoretical models with the polarimetric data on accretion discs around NSs and BHs.

\section*{Acknowledgments} 

This research (specifically in Sect 3.2) was supported by the Russian Science Foundation grant 20-12-00364. 
We also acknowledge support from the Jenny and Antti Wihuri foundation (VL) and the Academy of Finland grants 309308 (AV), 322779 and 333112 (JP). 

\bibliographystyle{aa}
\bibliography{references}

\begin{thebibliography}{49}
\expandafter\ifx\csname natexlab\endcsname\relax\def\natexlab#1{#1}\fi

\bibitem[{{Axelsson} \& {Veledina}(2021)}]{Axelsson21}
{Axelsson}, M. \& {Veledina}, A. 2021, \mnras [\eprint[arXiv]{2103.08795}]

\bibitem[{{Bambi} {et~al.}(2021){Bambi}, {Brenneman}, {Dauser}, {Garc{\'\i}a},
  {Grinberg}, {Ingram}, {Jiang}, {Liu}, {Lohfink}, {Marinucci}, {Mastroserio},
  {Middei}, {Nampalliwar}, {Nied{\'z}wiecki}, {Steiner}, {Tripathi}, \&
  {Zdziarski}}]{Bambi2021}
{Bambi}, C., {Brenneman}, L.~W., {Dauser}, T., {et~al.} 2021, \ssr, 217, 65

\bibitem[{{Bardeen} \& {Petterson}(1975)}]{BardeenPatterson1975}
{Bardeen}, J.~M. \& {Petterson}, J.~A. 1975, \apjl, 195, L65

\bibitem[{{Beloborodov}(2002)}]{B02}
{Beloborodov}, A.~M. 2002, \apjl, 566, L85

\bibitem[{{Bower} {et~al.}(2018){Bower}, {Broderick}, {Dexter}, {Doeleman},
  {Falcke}, {Fish}, {Johnson}, {Marrone}, {Moran}, {Moscibrodzka}, {Peck},
  {Plambeck}, \& {Rao}}]{Bower18}
{Bower}, G.~C., {Broderick}, A., {Dexter}, J., {et~al.} 2018, \apj, 868, 101

\bibitem[{{Chandrasekhar}(1960)}]{Cha60}
{Chandrasekhar}, S. 1960, {Radiative transfer} (New York: Dover)

\bibitem[{{Chandrasekhar} \& {Breen}(1947)}]{Chandrasekhar1947}
{Chandrasekhar}, S. \& {Breen}, F.~H. 1947, \apj, 105, 435

\bibitem[{{Chen} {et~al.}(1989){Chen}, {Halpern}, \& {Filippenko}}]{Chen89}
{Chen}, K., {Halpern}, J.~P., \& {Filippenko}, A.~V. 1989, \apj, 339, 742

\bibitem[{{Connors} {et~al.}(1980){Connors}, {Piran}, \& {Stark}}]{Connors1980}
{Connors}, P.~A., {Piran}, T., \& {Stark}, R.~F. 1980, \apj, 235, 224

\bibitem[{{Connors} \& {Stark}(1977)}]{ConnorsStark1977}
{Connors}, P.~A. \& {Stark}, R.~F. 1977, \nat, 269, 128

\bibitem[{{Dov{\v{c}}iak} {et~al.}(2008){Dov{\v{c}}iak}, {Muleri}, {Goosmann},
  {Karas}, \& {Matt}}]{Dovciak2008}
{Dov{\v{c}}iak}, M., {Muleri}, F., {Goosmann}, R.~W., {Karas}, V., \& {Matt},
  G. 2008, \mnras, 391, 32

\bibitem[{{Event Horizon Telescope Collaboration}
  {et~al.}(2021{\natexlab{a}}){Event Horizon Telescope Collaboration},
  {Akiyama}, {Algaba}, {Alberdi}, {Alef}, {Anantua}, {Asada}, {Azulay},
  {Baczko}, {Ball}, {Balokovi{\'c}}, {Barrett}, {Benson}, {Bintley},
  {Blackburn}, {Blundell}, {Boland}, {Bouman}, {Bower}, {Boyce}, {Bremer},
  {Brinkerink}, {Brissenden}, {Britzen}, {Broderick}, {Broguiere}, {Bronzwaer},
  {Byun}, {Carlstrom}, {Chael}, {Chan}, {Chatterjee}, {Chatterjee}, {Chen},
  {Chen}, {Chesler}, {Cho}, {Christian}, {Conway}, {Cordes}, {Crawford},
  {Crew}, {Cruz-Osorio}, {Cui}, {Davelaar}, {De Laurentis}, {Deane}, {Dempsey},
  {Desvignes}, {Dexter}, {Doeleman}, {Eatough}, {Falcke}, {Farah}, {Fish},
  {Fomalont}, {Ford}, {Fraga-Encinas}, {Freeman}, {Friberg}, {Fromm},
  {Fuentes}, {Galison}, {Gammie}, {Garc{\'\i}a}, {Gentaz}, {Georgiev}, {Goddi},
  {Gold}, {G{\'o}mez}, {G{\'o}mez-Ruiz}, {Gu}, {Gurwell}, {Hada}, {Haggard},
  {Hecht}, {Hesper}, {Ho}, {Ho}, {Honma}, {Huang}, {Huang}, {Hughes}, {Ikeda},
  {Inoue}, {Issaoun}, {James}, {Jannuzi}, {Janssen}, {Jeter}, {Jiang},
  {Jimenez-Rosales}, {Johnson}, {Jorstad}, {Jung}, {Karami}, {Karuppusamy},
  {Kawashima}, {Keating}, {Kettenis}, {Kim}, {Kim}, {Kim}, {Kim}, {Kino},
  {Koay}, {Kofuji}, {Koch}, {Koyama}, {Kramer}, {Kramer}, {Krichbaum}, {Kuo},
  {Lauer}, {Lee}, {Levis}, {Li}, {Li}, {Lindqvist}, {Lico}, {Lindahl}, {Liu},
  {Liu}, {Liuzzo}, {Lo}, {Lobanov}, {Loinard}, {Lonsdale}, {Lu}, {MacDonald},
  {Mao}, {Marchili}, {Markoff}, {Marrone}, {Marscher}, {Mart{\'\i}-Vidal},
  {Matsushita}, {Matthews}, {Medeiros}, {Menten}, {Mizuno}, {Mizuno}, {Moran},
  {Moriyama}, {Moscibrodzka}, {M{\"u}ller}, {Musoke}, {Mej{\'\i}as},
  {Michalik}, {Nadolski}, {Nagai}, {Nagar}, {Nakamura}, {Narayan}, {Narayanan},
  {Natarajan}, {Nathanail}, {Neilsen}, {Neri}, {Ni}, {Noutsos}, {Nowak},
  {Okino}, {Olivares}, {Ortiz-Le{\'o}n}, {Oyama}, {{\"O}zel}, {Palumbo},
  {Park}, {Patel}, {Pen}, {Pesce}, {Pi{\'e}tu}, {Plambeck}, {PopStefanija},
  {Porth}, {P{\"o}tzl}, {Prather}, {Preciado-L{\'o}pez}, {Psaltis}, {Pu},
  {Ramakrishnan}, {Rao}, {Rawlings}, {Raymond}, {Rezzolla}, {Ricarte},
  {Ripperda}, {Roelofs}, {Rogers}, {Ros}, {Rose}, {Roshanineshat}, {Rottmann},
  {Roy}, {Ruszczyk}, {Rygl}, {S{\'a}nchez}, {S{\'a}nchez-Arguelles}, {Sasada},
  {Savolainen}, {Schloerb}, {Schuster}, {Shao}, {Shen}, {Small}, {Sohn},
  {SooHoo}, {Sun}, {Tazaki}, {Tetarenko}, {Tiede}, {Tilanus}, {Titus}, {Toma},
  {Torne}, {Trent}, {Traianou}, {Trippe}, {van Bemmel}, {van Langevelde}, {van
  Rossum}, {Wagner}, {Ward-Thompson}, {Wardle}, {Weintroub}, {Wex}, {Wharton},
  {Wielgus}, {Wong}, {Wu}, {Yoon}, {Young}, {Young}, {Younsi}, {Yuan}, {Yuan},
  {Zensus}, {Zhao}, \& {Zhao}}]{EHT21a}
{Event Horizon Telescope Collaboration}, {Akiyama}, K., {Algaba}, J.~C.,
  {et~al.} 2021{\natexlab{a}}, \apjl, 910, L12

\bibitem[{{Event Horizon Telescope Collaboration}
  {et~al.}(2021{\natexlab{b}}){Event Horizon Telescope Collaboration},
  {Akiyama}, {Algaba}, {Alberdi}, {Alef}, {Anantua}, {Asada}, {Azulay},
  {Baczko}, {Ball}, {Balokovi{\'c}}, {Barrett}, {Benson}, {Bintley},
  {Blackburn}, {Blundell}, {Boland}, {Bouman}, {Bower}, {Boyce}, {Bremer},
  {Brinkerink}, {Brissenden}, {Britzen}, {Broderick}, {Broguiere}, {Bronzwaer},
  {Byun}, {Carlstrom}, {Chael}, {Chan}, {Chatterjee}, {Chatterjee}, {Chen},
  {Chen}, {Chesler}, {Cho}, {Christian}, {Conway}, {Cordes}, {Crawford},
  {Crew}, {Cruz-Osorio}, {Cui}, {Davelaar}, {De Laurentis}, {Deane}, {Dempsey},
  {Desvignes}, {Dexter}, {Doeleman}, {Eatough}, {Falcke}, {Farah}, {Fish},
  {Fomalont}, {Ford}, {Fraga-Encinas}, {Friberg}, {Fromm}, {Fuentes},
  {Galison}, {Gammie}, {Garc{\'\i}a}, {Gelles}, {Gentaz}, {Georgiev}, {Goddi},
  {Gold}, {G{\'o}mez}, {G{\'o}mez-Ruiz}, {Gu}, {Gurwell}, {Hada}, {Haggard},
  {Hecht}, {Hesper}, {Himwich}, {Ho}, {Ho}, {Honma}, {Huang}, {Huang},
  {Hughes}, {Ikeda}, {Inoue}, {Issaoun}, {James}, {Jannuzi}, {Janssen},
  {Jeter}, {Jiang}, {Jimenez-Rosales}, {Johnson}, {Jorstad}, {Jung}, {Karami},
  {Karuppusamy}, {Kawashima}, {Keating}, {Kettenis}, {Kim}, {Kim}, {Kim},
  {Kim}, {Kino}, {Koay}, {Kofuji}, {Koch}, {Koyama}, {Kramer}, {Kramer},
  {Krichbaum}, {Kuo}, {Lauer}, {Lee}, {Levis}, {Li}, {Li}, {Lindqvist}, {Lico},
  {Lindahl}, {Liu}, {Liu}, {Liuzzo}, {Lo}, {Lobanov}, {Loinard}, {Lonsdale},
  {Lu}, {MacDonald}, {Mao}, {Marchili}, {Markoff}, {Marrone}, {Marscher},
  {Mart{\'\i}-Vidal}, {Matsushita}, {Matthews}, {Medeiros}, {Menten}, {Mizuno},
  {Mizuno}, {Moran}, {Moriyama}, {Moscibrodzka}, {M{\"u}ller}, {Musoke}, {Mus
  Mej{\'\i}as}, {Michalik}, {Nadolski}, {Nagai}, {Nagar}, {Nakamura},
  {Narayan}, {Narayanan}, {Natarajan}, {Nathanail}, {Neilsen}, {Neri}, {Ni},
  {Noutsos}, {Nowak}, {Okino}, {Olivares}, {Ortiz-Le{\'o}n}, {Oyama},
  {{\"O}zel}, {Palumbo}, {Park}, {Patel}, {Pen}, {Pesce}, {Pi{\'e}tu},
  {Plambeck}, {PopStefanija}, {Porth}, {P{\"o}tzl}, {Prather},
  {Preciado-L{\'o}pez}, {Psaltis}, {Pu}, {Ramakrishnan}, {Rao}, {Rawlings},
  {Raymond}, {Rezzolla}, {Ricarte}, {Ripperda}, {Roelofs}, {Rogers}, {Ros},
  {Rose}, {Roshanineshat}, {Rottmann}, {Roy}, {Ruszczyk}, {Rygl},
  {S{\'a}nchez}, {S{\'a}nchez-Arguelles}, {Sasada}, {Savolainen}, {Schloerb},
  {Schuster}, {Shao}, {Shen}, {Small}, {Sohn}, {SooHoo}, {Sun}, {Tazaki},
  {Tetarenko}, {Tiede}, {Tilanus}, {Titus}, {Toma}, {Torne}, {Trent},
  {Traianou}, {Trippe}, {van Bemmel}, {van Langevelde}, {van Rossum}, {Wagner},
  {Ward-Thompson}, {Wardle}, {Weintroub}, {Wex}, {Wharton}, {Wielgus}, {Wong},
  {Wu}, {Yoon}, {Young}, {Young}, {Younsi}, {Yuan}, {Yuan}, {Zensus}, {Zhao},
  \& {Zhao}}]{EHT21b}
{Event Horizon Telescope Collaboration}, {Akiyama}, K., {Algaba}, J.~C.,
  {et~al.} 2021{\natexlab{b}}, \apjl, 910, L13

\bibitem[{{Gilfanov} {et~al.}(2003){Gilfanov}, {Revnivtsev}, \&
  {Molkov}}]{GRM03}
{Gilfanov}, M., {Revnivtsev}, M., \& {Molkov}, S. 2003, \aap, 410, 217

\bibitem[{{Gravity Collaboration} {et~al.}(2018){Gravity Collaboration},
  {Abuter}, {Amorim}, {Baub{\"o}ck}, {Berger}, {Bonnet}, {Brandner},
  {Cl{\'e}net}, {Coud{\'e} Du Foresto}, {de Zeeuw}, {Deen}, {Dexter}, {Duvert},
  {Eckart}, {Eisenhauer}, {F{\"o}rster Schreiber}, {Garcia}, {Gao}, {Gendron},
  {Genzel}, {Gillessen}, {Guajardo}, {Habibi}, {Haubois}, {Henning}, {Hippler},
  {Horrobin}, {Huber}, {Jim{\'e}nez-Rosales}, {Jocou}, {Kervella}, {Lacour},
  {Lapeyr{\`e}re}, {Lazareff}, {Le Bouquin}, {L{\'e}na}, {Lippa}, {Ott},
  {Panduro}, {Paumard}, {Perraut}, {Perrin}, {Pfuhl}, {Plewa}, {Rabien},
  {Rodr{\'\i}guez-Coira}, {Rousset}, {Sternberg}, {Straub}, {Straubmeier},
  {Sturm}, {Tacconi}, {Vincent}, {von Fellenberg}, {Waisberg}, {Widmann},
  {Wieprecht}, {Wiezorrek}, {Woillez}, \& {Yazici}}]{Gravity18}
{Gravity Collaboration}, {Abuter}, R., {Amorim}, A., {et~al.} 2018, \aap, 618,
  L10

\bibitem[{{Ingram} {et~al.}(2015){Ingram}, {Maccarone}, {Poutanen}, \&
  {Krawczynski}}]{Ingram15}
{Ingram}, A., {Maccarone}, T.~J., {Poutanen}, J., \& {Krawczynski}, H. 2015,
  \apj, 807, 53

\bibitem[{{Kerr}(1963)}]{Kerr1963}
{Kerr}, R.~P. 1963, \prl, 11, 237

\bibitem[{{Li} {et~al.}(2009){Li}, {Narayan}, \& {McClintock}}]{Li2009}
{Li}, L.-X., {Narayan}, R., \& {McClintock}, J.~E. 2009, \apj, 691, 847

\bibitem[{{Lightman} \& {Shapiro}(1975)}]{LightmanShapiro1975}
{Lightman}, A.~P. \& {Shapiro}, S.~L. 1975, \apjl, 198, L73

\bibitem[{{Loktev} {et~al.}(2020){Loktev}, {Salmi}, {N{\"a}ttil{\"a}}, \&
  {Poutanen}}]{Loktev20}
{Loktev}, V., {Salmi}, T., {N{\"a}ttil{\"a}}, J., \& {Poutanen}, J. 2020, \aap,
  643, A84

\bibitem[{{Loskutov} \& {Sobolev}(1979)}]{Losob79}
{Loskutov}, V.~M. \& {Sobolev}, V.~V. 1979, Astrofizika, 15, 241

\bibitem[{{Loskutov} \& {Sobolev}(1981)}]{LoskutovSobolev1981}
{Loskutov}, V.~M. \& {Sobolev}, V.~V. 1981, Astrofizika, 17, 97

\bibitem[{{Luminet}(1979)}]{Luminet1979}
{Luminet}, J.~P. 1979, \aap, 75, 228

\bibitem[{{Narayan} {et~al.}(2021){Narayan}, {Palumbo}, {Johnson}, {Gelles},
  {Himwich}, {Chang}, {Ricarte}, {Dexter}, {Gammie}, {Chael}, {Event Horizon
  Telescope Collaboration}, {Akiyama}, {Alberdi}, {Alef}, {Algaba}, {Anantua},
  {Asada}, {Azulay}, {Baczko}, {Ball}, {Balokovi{\'c}}, {Barrett}, {Benson},
  {Bintley}, {Blackburn}, {Blundell}, {Boland}, {Bouman}, {Bower}, {Boyce},
  {Bremer}, {Brinkerink}, {Brissenden}, {Britzen}, {Broderick}, {Broguiere},
  {Bronzwaer}, {Byun}, {Carlstrom}, {Chan}, {Chatterjee}, {Chatterjee}, {Chen},
  {Chen}, {Chesler}, {Cho}, {Christian}, {Conway}, {Cordes}, {Crawford},
  {Crew}, {Cruz-Osorio}, {Cui}, {Davelaar}, {De Laurentis}, {Deane}, {Dempsey},
  {Desvignes}, {Doeleman}, {Eatough}, {Falcke}, {Farah}, {Fish}, {Fomalont},
  {Ford}, {Fraga-Encinas}, {Friberg}, {Fromm}, {Fuentes}, {Galison},
  {Garc{\'\i}a}, {Gentaz}, {Georgiev}, {Goddi}, {Gold}, {G{\'o}mez},
  {G{\'o}mez-Ruiz}, {Gu}, {Gurwell}, {Hada}, {Haggard}, {Hecht}, {Hesper},
  {Ho}, {Ho}, {Honma}, {Huang}, {Huang}, {Hughes}, {Ikeda}, {Inoue}, {Issaoun},
  {James}, {Jannuzi}, {Janssen}, {Jeter}, {Jiang}, {Jimenez-Rosales},
  {Jorstad}, {Jung}, {Karami}, {Karuppusamy}, {Kawashima}, {Keating},
  {Kettenis}, {Kim}, {Kim}, {Kim}, {Kim}, {Kino}, {Koay}, {Kofuji}, {Koch},
  {Koyama}, {Kramer}, {Kramer}, {Krichbaum}, {Kuo}, {Lauer}, {Lee}, {Levis},
  {Li}, {Li}, {Lindqvist}, {Lico}, {Lindahl}, {Liu}, {Liu}, {Liuzzo}, {Lo},
  {Lobanov}, {Loinard}, {Lonsdale}, {Lu}, {MacDonald}, {Mao}, {Marchili},
  {Markoff}, {Marrone}, {Marscher}, {Mart{\'\i}-Vidal}, {Matsushita},
  {Matthews}, {Medeiros}, {Menten}, {Mizuno}, {Mizuno}, {Moran}, {Moriyama},
  {Moscibrodzka}, {M{\"u}ller}, {Musoke}, {Mej{\'\i}as}, {Nagai}, {Nagar},
  {Nakamura}, {Narayanan}, {Natarajan}, {Nathanail}, {Neilsen}, {Neri}, {Ni},
  {Noutsos}, {Nowak}, {Okino}, {Olivares}, {Ortiz-Le{\'o}n}, {Oyama},
  {{\"O}zel}, {Park}, {Patel}, {Pen}, {Pesce}, {Pi{\'e}tu}, {Plambeck},
  {PopStefanija}, {Porth}, {P{\"o}tzl}, {Prather}, {Preciado-L{\'o}pez},
  {Psaltis}, {Pu}, {Ramakrishnan}, {Rao}, {Rawlings}, {Raymond}, {Rezzolla},
  {Ripperda}, {Roelofs}, {Rogers}, {Ros}, {Rose}, {Roshanineshat}, {Rottmann},
  {Roy}, {Ruszczyk}, {Rygl}, {S{\'a}nchez}, {S{\'a}nchez-Arguelles}, {Sasada},
  {Savolainen}, {Schloerb}, {Schuster}, {Shao}, {Shen}, {Small}, {Sohn},
  {SooHoo}, {Sun}, {Tazaki}, {Tetarenko}, {Tiede}, {Tilanus}, {Titus}, {Toma},
  {Torne}, {Trent}, {Traianou}, {Trippe}, {van Bemmel}, {van Langevelde}, {van
  Rossum}, {Wagner}, {Ward-Thompson}, {Wardle}, {Weintroub}, {Wex}, {Wharton},
  {Wielgus}, {Wong}, {Wu}, {Yoon}, {Young}, {Young}, {Younsi}, {Yuan}, {Yuan},
  {Zensus}, {Zhao}, \& {Zhao}}]{Narayan21}
{Narayan}, R., {Palumbo}, D. C.~M., {Johnson}, M.~D., {et~al.} 2021, \apj, 912,
  35

\bibitem[{{Novikov} \& {Thorne}(1973)}]{NT73}
{Novikov}, I.~D. \& {Thorne}, K.~S. 1973, in Black Holes (Les Astres Occlus),
  ed. C.~{DeWitt} \& B.~{DeWitt} (New York: Gordon and Breach), 343--450

\bibitem[{{Pechenick} {et~al.}(1983){Pechenick}, {Ftaclas}, \& {Cohen}}]{PFC83}
{Pechenick}, K.~R., {Ftaclas}, C., \& {Cohen}, J.~M. 1983, \apj, 274, 846

\bibitem[{{Pineault}(1977)}]{Pineault1977pol_schw}
{Pineault}, S. 1977, \mnras, 179, 691

\bibitem[{{Pineault} \&
  {Roeder}(1977{\natexlab{a}})}]{PineaultRoeder1977kerr_num}
{Pineault}, S. \& {Roeder}, R.~C. 1977{\natexlab{a}}, \apj, 213, 548

\bibitem[{{Pineault} \&
  {Roeder}(1977{\natexlab{b}})}]{PineaultRoeder1977kerr_analyt}
{Pineault}, S. \& {Roeder}, R.~C. 1977{\natexlab{b}}, \apj, 212, 541

\bibitem[{{Poutanen}(2020{\natexlab{a}})}]{Poutanen2020bending}
{Poutanen}, J. 2020{\natexlab{a}}, \aap, 640, A24

\bibitem[{{Poutanen}(2020{\natexlab{b}})}]{Poutanen2020polarization}
{Poutanen}, J. 2020{\natexlab{b}}, \aap, 641, A166

\bibitem[{{Poutanen} \& {Beloborodov}(2006)}]{PB06}
{Poutanen}, J. \& {Beloborodov}, A.~M. 2006, \mnras, 373, 836

\bibitem[{{Poutanen} \& {Veledina}(2014)}]{PV14}
{Poutanen}, J. \& {Veledina}, A. 2014, \ssr, 183, 61

\bibitem[{{Rees}(1975)}]{Rees1975}
{Rees}, M.~J. 1975, \mnras, 171, 457

\bibitem[{{Revnivtsev} {et~al.}(1999){Revnivtsev}, {Gilfanov}, \&
  {Churazov}}]{Revnivtsev1999}
{Revnivtsev}, M., {Gilfanov}, M., \& {Churazov}, E. 1999, \aap, 347, L23

\bibitem[{{Reynolds}(2014)}]{Reynolds14}
{Reynolds}, C.~S. 2014, \ssr, 183, 277

\bibitem[{{Salmi} {et~al.}(2018){Salmi}, {N{\"a}ttil{\"a}}, \&
  {Poutanen}}]{SNP18}
{Salmi}, T., {N{\"a}ttil{\"a}}, J., \& {Poutanen}, J. 2018, \aap, 618, A161

\bibitem[{{Shakura} \& {Sunyaev}(1973)}]{SS73}
{Shakura}, N.~I. \& {Sunyaev}, R.~A. 1973, \aap, 24, 337

\bibitem[{{Shimura} \& {Takahara}(1995)}]{ST95}
{Shimura}, T. \& {Takahara}, F. 1995, \apj, 445, 780

\bibitem[{{Sobolev}(1949)}]{sob49}
{Sobolev}, V.~V. 1949, Uch. Zap. Leningrad Univ., 16

\bibitem[{{Sobolev}(1963)}]{Sob63}
{Sobolev}, V.~V. 1963, {A treatise on radiative transfer} (Princeton: Van
  Nostrand)

\bibitem[{{Stark} \& {Connors}(1977)}]{StarkConnors1977}
{Stark}, R.~F. \& {Connors}, P.~A. 1977, \nat, 266, 429

\bibitem[{{Suleimanov} {et~al.}(2020){Suleimanov}, {Poutanen}, \&
  {Werner}}]{SPW20}
{Suleimanov}, V.~F., {Poutanen}, J., \& {Werner}, K. 2020, \aap, 639, A33

\bibitem[{{Uttley} {et~al.}(2014){Uttley}, {Cackett}, {Fabian}, {Kara}, \&
  {Wilkins}}]{Uttley2014}
{Uttley}, P., {Cackett}, E.~M., {Fabian}, A.~C., {Kara}, E., \& {Wilkins},
  D.~R. 2014, \aapr, 22, 72

\bibitem[{{Viironen} \& {Poutanen}(2004)}]{VP04}
{Viironen}, K. \& {Poutanen}, J. 2004, \aap, 426, 985

\bibitem[{{Walker} \& {Penrose}(1970)}]{WalkerPenrose1970}
{Walker}, M. \& {Penrose}, R. 1970, Communications in Mathematical Physics, 18,
  265

\bibitem[{{Weisskopf} {et~al.}(2016){Weisskopf}, {Ramsey}, {O'Dell}, {Tennant},
  {Elsner}, {Soffitta}, {Bellazzini}, {Costa}, {Kolodziejczak}, {Kaspi},
  {Muleri}, {Marshall}, {Matt}, \& {Romani}}]{weisskopf16}
{Weisskopf}, M.~C., {Ramsey}, B., {O'Dell}, S., {et~al.} 2016, in Society of
  Photo-Optical Instrumentation Engineers (SPIE) Conference Series, Vol. 9905,
  Space Telescopes and Instrumentation 2016: Ultraviolet to Gamma Ray, ed.
  J.-W.~A. {den Herder}, T.~{Takahashi}, \& M.~{Bautz}, 990517

\bibitem[{{Yuan} \& {Narayan}(2014)}]{YN14}
{Yuan}, F. \& {Narayan}, R. 2014, \araa, 52, 529

\bibitem[{{Zhang} {et~al.}(2019){Zhang}, {Santangelo}, {Feroci}, {Xu}, {Lu},
  {Chen}, {Feng}, {Zhang}, {Brandt}, {Hernanz}, {Baldini}, {Bozzo}, {Campana},
  {De Rosa}, {Dong}, {Evangelista}, {Karas}, {Meidinger}, {Meuris}, {Nandra},
  {Pan}, {Pareschi}, {Orleanski}, {Huang}, {Schanne}, {Sironi}, {Spiga},
  {Svoboda}, {Tagliaferri}, {Tenzer}, {Vacchi}, {Zane}, {Walton}, {Wang},
  {Winter}, {Wu}, {in't Zand}, {Ahangarianabhari}, {Ambrosi}, {Ambrosino},
  {Barbera}, {Basso}, {Bayer}, {Bellazzini}, {Bellutti}, {Bertucci},
  {Bertuccio}, {Borghi}, {Cao}, {Cadoux}, {Campana}, {Ceraudo}, {Chen}, {Chen},
  {Chevenez}, {Civitani}, {Cui}, {Cui}, {Dauser}, {Del Monte}, {Di Cosimo},
  {Diebold}, {Doroshenko}, {Dovciak}, {Du}, {Ducci}, {Fan}, {Favre},
  {Fuschino}, {G{\'a}lvez}, {Gao}, {Ge}, {Gevin}, {Grassi}, {Gu}, {Gu}, {Han},
  {Hong}, {Hu}, {Ji}, {Jia}, {Jiang}, {Kennedy}, {Kreykenbohm}, {Kuvvetli},
  {Labanti}, {Latronico}, {Li}, {Li}, {Li}, {Li}, {Li}, {Limousin}, {Liu},
  {Liu}, {Lu}, {Luo}, {Macera}, {Malcovati}, {Martindale}, {Michalska}, {Meng},
  {Minuti}, {Morbidini}, {Muleri}, {Paltani}, {Perinati}, {Picciotto},
  {Piemonte}, {Qu}, {Rachevski}, {Rashevskaya}, {Rodriguez}, {Schanz}, {Shen},
  {Sheng}, {Song}, {Song}, {Sgro}, {Sun}, {Tan}, {Uttley}, {Wang}, {Wang},
  {Wang}, {Wang}, {Wang}, {Wang}, {Watts}, {Wen}, {Wilms}, {Xiong}, {Yang},
  {Yang}, {Yang}, {Yu}, {Zhang}, {Zampa}, {Zampa}, {Zdziarski}, {Zhang},
  {Zhang}, {Zhang}, {Zhang}, {Zhang}, {Zhang}, {Zhang}, {Zhang}, {Zhao},
  {Zheng}, {Zhou}, {Zorzi}, \& {Zwart}}]{Zhang2019eXTP}
{Zhang}, S., {Santangelo}, A., {Feroci}, M., {et~al.} 2019, Science China
  Physics, Mechanics, and Astronomy, 62, 29502

\end{thebibliography}

\appendix 

\section{Rotation of polarization plane}
\label{sec:deriv}

\begin{figure}
\centering
\includegraphics[width=0.85\linewidth]{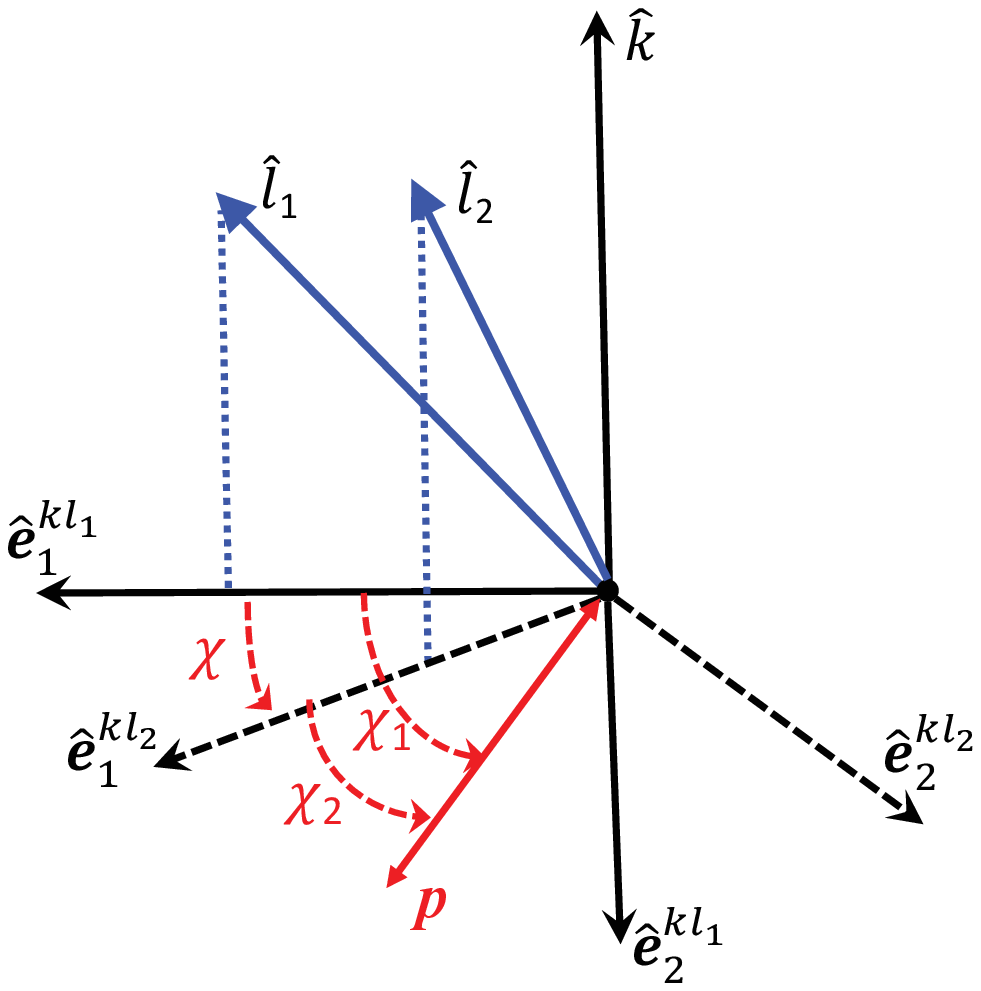}
\caption{Polarization bases $\mathcal{B}(\unit{k}, \bm{l}_1)$ and $\mathcal{B}(\unit{k}, \bm{l}_2)$, polarization vector $\bm{p}$ and corresponding PAs.  
}
\label{fig:twopolbases}
\end{figure}

\subsection{Rotation of PA for two polarization bases}

For a photon moving along a unit vector $\unit{k}$ one can define a polarization  basis of two unit vectors perpendicular to each other as well as to $\unit{k}$ as
\begin{equation}
\label{eq:base0n}
\mathcal{B}(\unit{k}, \bm{l}) = 
\left\{
\unit{e}_1^{k,l},\, 
\unit{e}_2^{k,l}
\right\}  
=  \left\{
\frac{\bm{l}-(\unit{k}\cdot\bm{l}) \unit{k}}{|\unit{k} \times \bm{l}|},\; 
\frac{\unit{k} \times \bm{l}}{|\unit{k} \times \bm{l}|}
\right\} ,
\end{equation} 
where $\bm{l}$ is an arbitrary vector not collinear with $\unit{k}$. 
The angle the polarization plane of such a photon makes with the vector $\unit{e}_1^{k,l}$ measured counterclockwise as viewed by the observer defines the polarization angle (PA) of the photon in this basis.

Two bases, $\mathcal{B}(\unit{k}, \bm{l}_1)$ and $\mathcal{B}(\unit{k}, \bm{l}_2)$, built around the same photon vector $\unit{k}$, differ only by rotation around $\unit{k}$ by an angle $\chi(\unit{k}, \bm{l}_2 \rightarrow \bm{l}_1)$ (see Fig.~\ref{fig:twopolbases}) that may be defined by trigonometric functions as 
\begin{eqnarray}
\cos \chi(\unit{k}, \bm{l}_2 \rightarrow \bm{l}_1) &=& \unit{e}_2^{k,l_1}\cdot\unit{e}_2^{k,l_2}=\unit{e}_1^{k,l_1}\cdot\unit{e}_1^{k,l_2}, \\
\sin \chi(\unit{k}, \bm{l}_2 \rightarrow \bm{l}_1) &=& \unit{e}_2^{k,l_1}\cdot\unit{e}_1^{k,l_2}=-\unit{e}_1^{k,l_1}\cdot\unit{e}_2^{k,l_2}.
\end{eqnarray}
In this case, the PA (i.e. position angle of electric vector of a photon moving along $\unit{k}$) in basis 1 can be computed from the PA in basis 2 as $\chi_1=\chi_2+\chi$. 
The angle $\chi$ can be expressed through its tangent 
\begin{eqnarray} 
 \label{eq:orthotan}
\tan{\chi}(\unit{k}, \bm{l}_2 \rightarrow \bm{l}_1) &= & \frac{\unit{e}_2^{k,l_1}\cdot\unit{e}_1^{k,l_2}}{ \unit{e}_2^{k,l_1}\cdot\unit{e}_2^{k,l_2}}
=\frac{(\unit{k} \times \bm{l}_1)\cdot\left(\bm{l}_2-(\unit{k}\cdot\bm{l}_2) \unit{k}\right)}{(\unit{k} \times \bm{l}_1)\cdot(\unit{k} \times \bm{l}_2)}
 \nonumber \\
&= & \frac{\unit{k} \cdot (\bm{l}_1 \cross \bm{l}_2)  }{
\bm{l}_1 \cdot \bm{l}_2 - (\unit{k} \cdot \bm{l}_1)( \unit{k} \cdot \bm{l}_2 )} . 
\end{eqnarray}

\subsection{Rotation of PA due to light bending}
\label{sec:app_gr} 

Let us now quantify the effect of the gravitational light bending only.  
Let us assume that the Stokes vector of emitted radiation is defined at the disc surface in the lab (non-rotating) frame in the basis related to the disc normal $\unit{n}$, i.e. $\mathcal{B}(\unit{k}_0, \unit{n})$. 
On the other hand, the observer defines PA in the basis formed by $\unit{n}$ and the photon momentum at infinity $\unit{o}$, i.e. $\mathcal{B}(\unit{o}, \unit{n})$. 
The way to find the rotation of polarization plane between those two bases is to note that photon trajectories are flat in the Schwarzschild metric and the polarization vector is parallel transported along trajectory.  
Since the photon trajectory plane is defined by the vectors $\unit{r}$ and $\unit{o}$ (as well as $\unit{k}_0$), the PA measured in bases $\mathcal{B}(\unit{o}, \unit{r})$ and $\mathcal{B}(\unit{k}_0, \unit{r})$ is the same. 
Thus, the GR effect on PA consists of two terms only. 
The first one is rotation of polarization vector due to transformation from the basis $\mathcal{B}(\unit{k}_0, \unit{n})$ to $\mathcal{B}(\unit{k}_0, \unit{r})$, i.e. $\chi(\unit{k}_0,\unit{n} \rightarrow \unit{r})$. 
The second one is the rotation angle  $\chi(\unit{o},\unit{r} \rightarrow \unit{n})$ from basis 
$\mathcal{B}(\unit{o}, \unit{r})$ to $\mathcal{B}(\unit{o}, \unit{n})$. 
The total effect is then 
\begin{equation}
\label{eq:chiGR}
\chi^{\text{GR}} = 
\chi(\unit{k}_0,\unit{n} \rightarrow \unit{r}) +\chi(\unit{o},\unit{r} \rightarrow \unit{n}) , 
\end{equation}
where 
\begin{eqnarray}
\tan \chi(\unit{k}_0,\unit{n} \rightarrow \unit{r}) 
&= &
 \frac{\unit{k}_0\cdot (\unit{n} \cross \unit{r})  }{
 (\unit{k}_0 \cdot \unit{r})( \unit{k}_0 \cdot \unit{n} )} = -\frac{\tan i\ \sin\varphi}{\cos\alpha}, \\
\tan \chi(\unit{o},\unit{r} \rightarrow \unit{n}) 
&= &
\frac{\unit{o} \cdot (\unit{r} \cross \unit{n})  }{
(\unit{o} \cdot \unit{n})( \unit{o} \cdot \unit{r} )} = \frac{\tan i\ \sin\varphi}{\cos\psi},
\end{eqnarray}
where we used the facts that $\unit{n} \cross \unit{r}=\unit{\varphi}$, \textbf{$\unit{r}\cdot \unit{\varphi}=0$} and $\unit{r}\cdot \unit{n}=0$, with the rest of scalar products obtained directly from  Eqs.~\eqref{eq:geom_vectors}--\eqref{eq:circvel}:  
\begin{eqnarray}
\unit{o}\cdot \unit{r}& = & \cos \psi, \\  
\unit{k}_0\cdot \unit{r}& = &  \cos\alpha  , \\
\unit{o}\cdot \unit{n}& = & \cos i, \\  
\unit{k}_0\cdot \unit{n}& = & \frac{ \sin \alpha}{\sin\psi} \cos i = \frac{ \sin \alpha}{\sin\psi} \unit{o}\cdot \unit{n}  , \\
\unit{o}\cdot \unit{\varphi}& = & -  \sin i\ \sin\varphi, \\  
\unit{k}_0\cdot \unit{\varphi}& = & - \frac{ \sin \alpha}{\sin\psi} \sin i\ \sin\varphi = \frac{ \sin \alpha}{\sin\psi}  \unit{o}\cdot \unit{\varphi}. 
\end{eqnarray}
We thus get the tangent of the sum of the two rotation angles 
\begin{eqnarray}
\label{eq:tanGR_der}
\tan\chi^{\text{GR}}&=& \tan\left(\chi(\unit{k}_0,\unit{n} \rightarrow \unit{r}) +\chi(\unit{o},\unit{r} \rightarrow \unit{n}) \right) \nonumber\\ 
&=&-\frac{(\cos\alpha-\cos\psi)(\unit{o} \cdot \unit{\varphi})(\unit{o} \cdot \unit{n})}
{\cos\alpha \cos\psi (\unit{o} \cdot \unit{n})^2 + (\unit{o} \cdot \unit{\varphi})^2 } 
 \nonumber \\ 
&=&\frac{ -a(\unit{o} \cdot \unit{\varphi})(\unit{o} \cdot \unit{n})}
{1-(\unit{o} \cdot \unit{n})^2  + a (\unit{o}\cdot\unit{r}) } \nonumber \\ 
&=&\frac{a\cos i\sin\varphi}{\sin i + a\cos\varphi} , 
\end{eqnarray}
where 
\begin{equation}
a= \frac{\cos\alpha-\cos\psi}{1-\cos\alpha\cos\psi}
\end{equation}
and we used the relation between directional cosines 
\begin{equation}
(\unit{o} \cdot \unit{n})^2+
(\unit{o}\cdot\unit{r})^2 +
(\unit{o}\cdot\unit{\varphi})^2=1.
\end{equation}

We see that $\chi^{\text{GR}}=0$ at $\varphi=0$ where it changes the sign. 
Using \citet{B02} approximation for the light bending angle, $\cos\alpha=u + (1-u)\cos\psi$, we get 
\begin{equation}
a^{-1}=r+(r-1)\cos\psi, 
\end{equation}
which allows us to obtain a simple approximate expression for the rotation angle due to the light bending only: 
\begin{equation}\label{eq:tanGR_B02}
\tan\chi^{\text{GR}}_{\rm B02} \approx 
\frac{\cos i\sin\varphi}{r\sin i + (r \sin^2 i +\cos^2 i) \,\cos\varphi}  .
\end{equation}
The extrema are reached at 
\begin{equation}\label{eq:cosphiGR_ext}
\cos\varphi_{\rm ext} = - \frac{r \sin^2 i +\cos^2 i}{r\sin i} .
\end{equation}
This approximation works whenever the absolute value of the right-hand-side of Eq.~\eqref{eq:cosphiGR_ext} is smaller than unity, e.g. for $r=3$ the inclination should be $i>30\degr$.
For example, for $r=3$ and $i=60\degr$, the extrema appear at 
$\cos\varphi_{\rm ext,B02}=-5/(3\sqrt{3})$, i.e. at $\varphi_{\rm ext,B02}\approx180\degr\mp 16\degr$, corresponding to $\tan\chi^{\text{GR}}_{\rm ext,B02}=1/\sqrt{2}$ (or $\chi^{\text{GR}}_{\rm ext,B02}=\pm35\fdg3$), while the exact calculations give $\varphi_{\rm ext}\approx 180\degr\mp 17\fdg6$ and $\chi^{\text{GR}}_{\rm ext}\approx\pm31\fdg9$.

\subsection{Rotation of PA due to aberration}
\label{sec:app_sr} 

Now, let us consider effect of the special relativity (SR) on the PA. 
In the local frame we assume the element of the disc symmetric with respect to the local normal vector to the disc surface $\unit{n}$. Therefore the properties of the outgoing radiation in the local frame are only defined by the angle it makes with the vertical direction. 
An atmosphere model under such an assumption would give the polarization vector in $\mathcal{B}(\unit{k}_0', \unit{n})$ to be collinear to one of the basis vectors. 
The rotation of PA due to aberration can be decomposed into three rotations. 
First, from the basis $\mathcal{B}(\unit{k}_0', \unit{n})$ to $\mathcal{B}(\unit{k}_0', \unit{v})$. 
Second, from $\mathcal{B}(\unit{k}_0', \unit{v})$ to $\mathcal{B}(\unit{k}_0, \unit{v})$. 
Third, from $\mathcal{B}(\unit{k}_0, \unit{v})$ to $\mathcal{B}(\unit{k}_0, \unit{n})$. 
We note that the second rotation is actually zero, because three vectors  $\unit{k}_0',\unit{k}_0$ and $\unit{v}$ lie in the same plane, with the photon momenta unit vectors connected via the Lorenz transformation \eqref{eq:k0Lorentz}. 
Thus the total rotation of the PA due to the SR may be written as
\begin{equation}
\label{eq:chiSR}
\chi^{\text{SR}} = 
\chi(\unit{k}_0',\unit{n} \rightarrow \unit{v}) +\chi(\unit{k}_0,\unit{v} \rightarrow \unit{n}) .
\end{equation}
Noting that $\unit{v} \cdot \unit{n}=0$, each angle can be computed using Eq.~\eqref{eq:orthotan}: 
\begin{eqnarray} \label{eq:chiSR1}
\tan{\chi(\unit{k}_0',\unit{n} \rightarrow \unit{v}}) &= & 
\frac{\unit{k}_0' \cdot (\unit{n} \cross \unit{v})  }{
      (\unit{k}_0' \cdot \unit{n})(\unit{k}_0' \cdot \unit{v}) } ,\\
\label{eq:chiSR2} \tan{\chi(\unit{k}_0,\unit{v} \rightarrow \unit{n}}) &= & 
\frac{\unit{k}_0 \cdot (\unit{v} \cross \unit{n})  }{
      (\unit{k}_0 \cdot \unit{v})(\unit{k}_0 \cdot \unit{n}) }  .
\end{eqnarray}
Using the relation $\unit{v} \cross \unit{n}=\unit{r}$ valid in our case, and expression \eqref{eq:k0Lorentz} for the Lorentz transformation, we get 
\begin{eqnarray}
\frac{ \unit{k}_0'\cdot \unit{n} }{\unit{k}_0\cdot \unit{n}} &=& 
\frac{ \unit{k}_0'\cdot \unit{r} }{\unit{k}_0\cdot \unit{r}} = 
\delta = \frac{1}{\gamma(1-\beta\unit{k}_0 \cdot \unit{v})} , \\
\unit{k}_0'\cdot \unit{v} &=&
\frac{ \unit{k}_0 \cdot \unit{v} -\beta}{1 - \beta \unit{k}_0 \cdot \unit{v}} .
\end{eqnarray}
The tangent of the sum of two angles is then 
\begin{eqnarray} \label{eq:tanSR}
&& \tan\chi^{\text{SR}} =
\frac{  ( \unit{k}_0 \cdot \unit{r} ) (\unit{k}_0' \cdot \unit{n})(\unit{k}_0' \cdot \unit{v} )  -(\unit{k}_0' \cdot \unit{r}) (\unit{k}_0 \cdot \unit{v} )(\unit{k}_0 \cdot \unit{n}) }{
    (\unit{k}_0' \cdot \unit{n} ) (\unit{k}_0' \cdot \unit{v} ) (\unit{k}_0 \cdot \unit{n}) (\unit{k}_0 \cdot \unit{v} ) + (\unit{k}_0' \cdot \unit{r} ) (\unit{k}_0 \cdot \unit{r}) }    \nonumber \\
&=& 
\frac{( \unit{k}_0 \cdot \unit{r} ) (\unit{k}_0 \cdot \unit{n})  \left[ (\unit{k}_0 \cdot \unit{v} - \beta)  - (1-\beta\unit{k}_0\cdot\unit{v})  (\unit{k}_0 \cdot \unit{v} ) \right] }{
     (\unit{k}_0 \cdot \unit{v} - \beta)  (\unit{k}_0 \cdot \unit{v} )(\unit{k}_0 \cdot \unit{n})^2 + (1-\beta\unit{k}_0\cdot\unit{v}) (\unit{k}_0 \cdot \unit{r})^2 } \nonumber \\
&=&
\frac{  \beta(\unit{k}_0 \cdot \unit{r})(\unit{k}_0 \cdot \unit{n}) \left((\unit{k}_0 \cdot \unit{v})^2-1\right)   }{
      (\unit{k}_0 \cdot \unit{v} )^2(\unit{k}_0 \cdot \unit{n})^2 + (\unit{k}_0 \cdot \unit{r})^2 -\beta (\unit{k}_0\cdot\unit{v}) \left[ (\unit{k}_0 \cdot \unit{r})^2+(\unit{k}_0 \cdot \unit{n})^2\right] }  \nonumber \\
&=&- \frac{\beta(\unit{k}_0 \cdot \unit{r})(\unit{k}_0 \cdot \unit{n})}{
      1-(\unit{k}_0 \cdot \unit{n})^2  -\beta (\unit{k}_0\cdot\unit{v}) }  
      = -\beta\ \frac{ \cos\alpha  \cos\zeta  }{\sin^2\zeta -\beta \cos\xi } ,
\end{eqnarray}
where we used the relation for the directional cosines
\begin{equation}
   (\unit{k}_0 \cdot \unit{r})^2 + (\unit{k}_0 \cdot \unit{n})^2 + (\unit{k}_0 \cdot \unit{v})^2 =1 . 
\end{equation}
In flat space, we substitute $\alpha=\psi$, $\zeta=i$, $\cos\xi=-\sin i\sin\varphi$ and obtain Eq.~\eqref{eq:tanSRflat}. 

In a more general case, when vectors $\unit{n}, \unit{r}$, and $\unit{v}$ do not form the orthonormal basis, the expression for $\tan\chi^{\text{SR}}$ becomes somewhat more cumbersome. 
However, if the velocity is perpendicular to the normal vector (with relation to which we measure the PA in the first place), i.e. $\unit{v} \cdot \unit{n}=0$, then the rotation angle can still be written in a manner similar to the penultimate expression of Eq.~\eqref{eq:tanSR}:
\begin{equation}\label{eq:tanSRgeneral}
\tan\chi^{\text{SR}} = - \frac{\beta(\unit{k}_0 \cdot  (\unit{v} \cross \unit{n}))(\unit{k}_0 \cdot \unit{n})}{
      1-(\unit{k}_0 \cdot \unit{n})^2  -\beta (\unit{k}_0\cdot\unit{v}) } . 
\end{equation}

\end{document}